\documentclass[aps,
footinbib,
superscriptaddress,
showpacs,twocolumn,
prl]{revtex4-1}
\usepackage{moreverb}
\usepackage{graphicx} %Include figure files
\usepackage{color}
\usepackage{cancel}
\usepackage{amssymb}
\usepackage{amsmath}
\usepackage{amsfonts}
\usepackage{epstopdf}
\begin{document}
\title{Dynamics of interacting fermions in spin-dependent potentials}
\author{Andrew P. Koller}
\thanks{A.P.K. and M.L.W. contributed equally to this work.}
\affiliation{Department of Physics, University of Colorado,
Boulder, CO 80309}
\affiliation{JILA, NIST, Center for Theory of Quantum Matter, University of Colorado,
Boulder, CO 80309}
\author{Michael L. Wall}
\thanks{A.P.K. and M.L.W. contributed equally to this work.}
\affiliation{JILA, NIST, Center for Theory of Quantum Matter, University of Colorado,
Boulder, CO 80309}
\author{Josh Mundinger}
\affiliation{Department of Mathematics and Statistics, Swarthmore College, 500 College Avenue, Swarthmore, PA 19081}
\author{Ana Maria Rey}
\affiliation{Department of Physics, University of Colorado,
Boulder, CO 80309}
\affiliation{JILA, NIST, Center for Theory of Quantum Matter, University of Colorado,
Boulder, CO 80309}
\begin{abstract}
Recent experiments with dilute trapped Fermi gases observed that weak interactions can drastically modify spin transport dynamics and give rise to robust collective effects including global demagnetization, macroscopic spin waves, spin segregation, and spin self-rephasing. In this work we develop a framework for studying the dynamics of weakly interacting fermionic gases following a spin-dependent change of the trapping potential which illuminates the interplay between spin, motion, Fermi statistics, and interactions. The key idea is the projection of the state of the system onto a set of lattice spin models defined on the single-particle mode space. Collective phenomena, including the global spreading of quantum correlations in real space, arise as a consequence of the long-ranged character of the spin model couplings. This approach achieves good agreement with prior measurements and suggests a number of directions for future experiments. 
\end{abstract}

\maketitle

The interplay between spin and motional degrees of freedom in interacting electron systems has been a long-standing research topic in condensed matter physics. Interactions can modify the behavior of individual electrons  and give rise to emergent collective phenomena such as superconductivity and colossal magnetoresistance~\cite{CMR}. Theoretical understanding  of non-equilibrium dynamics in interacting fermionic matter is limited, however, and many open questions remain. Ultracold atomic Fermi gases, with precisely controllable parameters, offer an outstanding opportunity to investigate the emergence of collective behavior in out-of-equilibrium settings.

\begin{figure} %Retain asterik for wide figure and wide figure captions
\includegraphics[scale=.38]{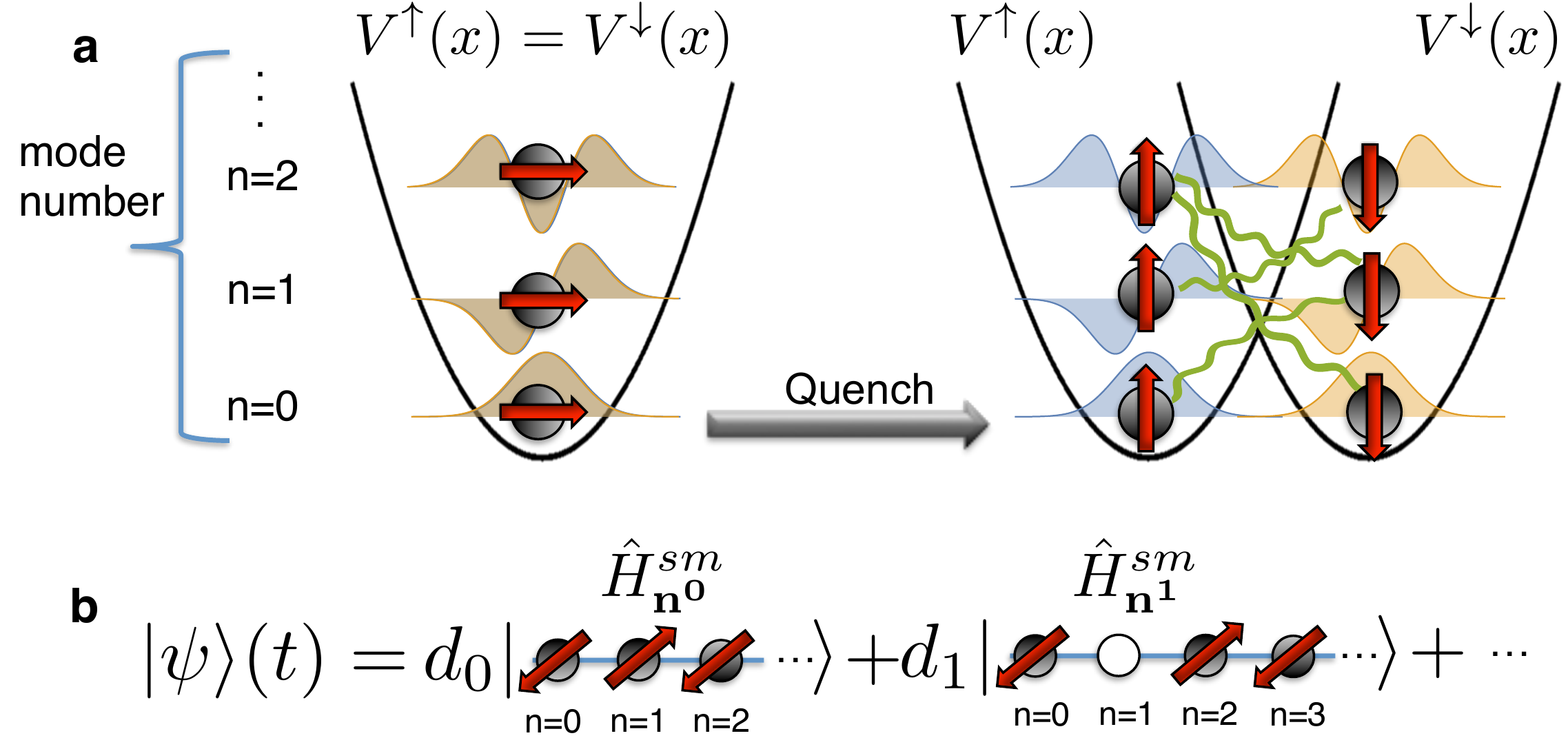}
\caption
{ \label{diagram}
\raggedright
(Color online) (a) Atoms spin-polarized along $X$ occupy single-particle eigenstates, labeled by mode number $n$. The potential is quenched to a spin-dependent form, and dynamics result from a spin model with long ranged interactions (green wavy lines) in energy space. (b) The state $|\psi \rangle$ is a coherent superposition of spins in many mode configurations (unoccupied modes are represented by open circles). In each configuration particles are localized in mode space, with spin model Hamiltonian $\hat H^{sm}_i$. Coherences between the configurations capture motional effects.}
\end{figure}

Progress in this direction has been made in recent experiments with ultracold spin-1/2 fermionic vapors, where initially spin-polarized gases were subjected to a spin-dependent trapping potential (Fig.~\ref{diagram}) implemented by a magnetic field gradient \cite{Koschorreck2013, Bardon14, Trotzky15}, or a spin-dependent harmonic trapping frequency \cite{Deutsch2010, Lewandowski2002, Du2008, Du2009} -- equivalent to a spatially-varying gradient. Even in the weakly interacting regime, drastic modifications of the single-particle dynamics were reported. Moreover, despite the local character of the interactions, collective phenomena were observed, including demagnetization and transverse spin-waves in the former, and a time-dependent separation (segregation) of the spin densities and spin self-rephasing in the latter. Although mean-field and kinetic theory formulations have explained some of these phenomena~\cite{Fuchs2002, Williams2002, Bradley2002,Du2009, Natu2009, ebling11, Bruun2011, Piechon2009, xu2015, Goulko2013,Enss2015}, a theory capable of describing all the time scales and the interplay between spin, motion, and interactions has not been developed. 

In this work, we develop a framework that accounts for the coupling of spin and motion in weakly interacting Fermi gases. We qualitatively reproduce and explain all phenomena of the aforementioned experiments and obtain quantitative agreement with the results of Ref.~\cite{Du2008}. In this formulation the state of the system is represented as a superposition of spin configurations which live on lattices whose sites correspond to modes of the underlying single-particle system. Within each configuration, the dynamics is described by a spin model with long-ranged couplings which generates collective quantum correlations and entanglement. Each sector evolves independently and the accumulated phase differences between sectors capture the interplay of spin and motion (Fig.~\ref{diagram} b). Using this formulation, we gain a great deal of insight
about the dynamics, and can extract analytic solutions
for spin observables and correlations in several limits. Although spin models in energy space~\cite{Oktel2002, Gibble2009, Rey2009, Yu2010, Hazlett2013, Koller14,Beverland14} have been used before and agreed well with experiments~\cite{Swallows2011,Deutsch2010,Maineult2012,Martin2013,Hazlett2013,Pechkis2013,Yan2013}, their use was limited to pure spin dynamics (no motion). Our formulation allows us to track motional degrees of freedom, compute local observables, and determine how correlations spread in real space. This opens a route for investigations of generic interacting spin-motion coupled systems beyond current capabilities. Our predictions also suggest directions for future experiments in the weakly interacting regime, which might, for instance, investigate the collective rise of quantum correlations. In contrast to strongly coupled ultracold gases, where motion is quickly suppressed and features of the dynamics tend to be universal \cite{Sommer2011,Koschorreck2013,Makotyn2014}, in the weakly-interacting regime spin, motion, and interactions are all important and must be treated on the same level.

A wide variety of analytical and numerical tools have been developed for lattice quantum spin models \cite{Schollwoeck,Polkovnikov2010, Schachenmayer2015, Pucci2015,Emch1966, Radin1970, Kastner2011, Foss-Feig2013}, making a spin model description of fermions potentially very useful. To demonstrate the capabilities of this approach, we use time-dependent matrix product state methods which are efficient in one-dimension~\footnote{The matrix product state studies of the main text were performed using
extensions of the open source MPS library~\cite {OSMPS,Wall_Carr_12}, and are described further in the supplement~\cite {supplement}.}. 
 
%Numerical simulations of non-equilibrium fermionic matter are notoriously difficult, and for many situations no efficient algorithms presently exist. In contrast, a wide variety of powerful analytical and numerical tools have been developed for lattice quantum spin models \cite{Schollwoeck,Polkovnikov2010, Schachenmayer2015, Pucci2015,Emch1966, Radin1970, Kastner2011, Foss-Feig2013}, making a spin model description of fermionic systems potentially very useful. To demonstrate the capabilities of this approach, we use time-dependent matrix product state methods which are efficient in one-dimension~\footnote{The matrix product state studies of the main text were performed using
%  extensions of the open source MPS library~\cite {OSMPS,Wall_Carr_12}, and are
%  described further in the supplement~\cite {supplement}.}. We simulate systems of $N=10-20$ particles; due to the coupling of spin and motion, the complexity of these simulations is similar to that of long-ranged and inhomogenous pure spin systems with $N\sim100$ spins. We emphasize, however, that the mapping to a spin model, the corresponding analytic solutions, and the physical interpretations are valid in arbitrary dimensions. Thus the method described here will be useful for more generic cases as numerical techniques able to handle larger spin systems continue to improve.  
\begin{figure*} %Retain asterik for wide figure and wide figure captions
\includegraphics[width=2.0\columnwidth]{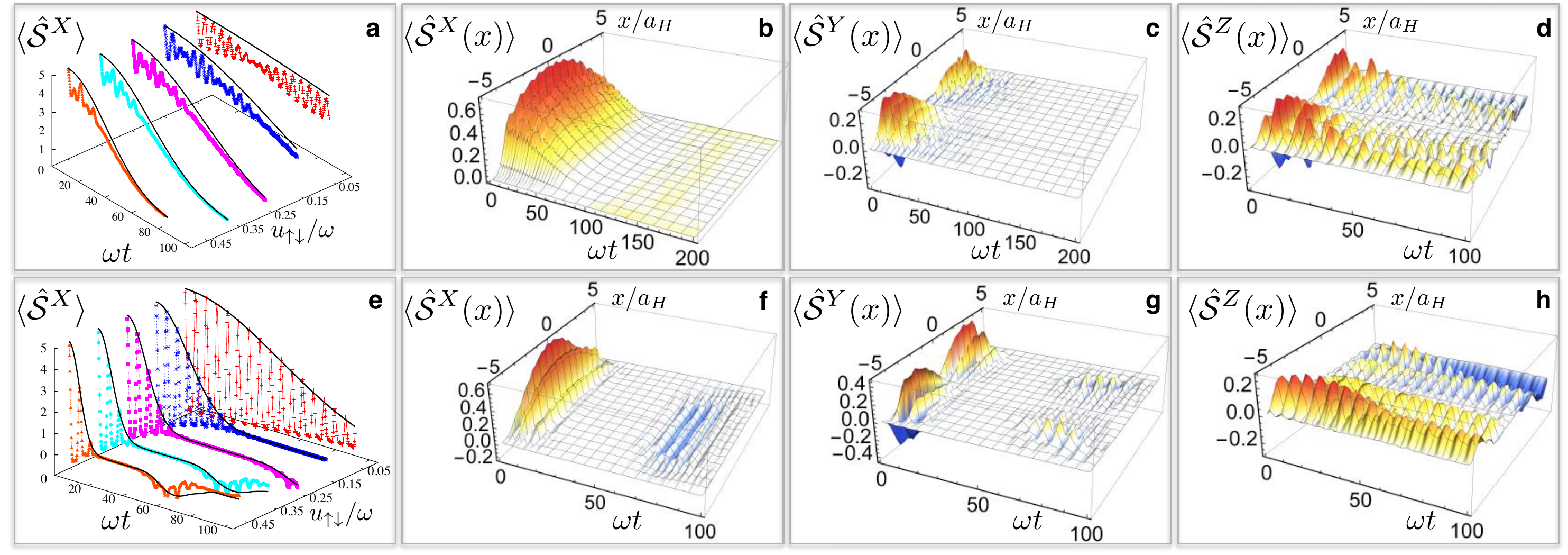}
\caption
{ \label{constant_fig}
\raggedright
(Color online) Magnetization dynamics for a constant gradient. Collective $\langle \hat{ \mathcal S}^X \rangle$ for a $x_0=0.1a_H$ (a) (and $x_0=0.3a_H$ (e)) displays global interaction-induced demagnetization, which damps single-particle oscillations. Collective (generic) Ising solutions, black lines, give the demagnetization envelopes. Local magnetizations $\langle \hat{\mathcal S}^{X,Y,Z}(x) \rangle$ with $x_0=0.1a_H$ (b-d) (and $x_0=0.3a_H$ f-h) reflect similar behavior, both shown with $u_{\uparrow\downarrow}=0.35\omega$.}
\end{figure*}

{ \it Setup--} We consider $N$ identical fermionic atoms of mass $m_a$ with a spin-1/2 degree of freedom $\alpha\in\{\uparrow,\downarrow\}$ trapped in a one dimensional harmonic oscillator of frequency $\omega$, $V^0(x)=\frac{1}{2}m_a\omega^2 x^2$. The gas begins spin-polarized in the $\downarrow$ state and atoms populate distinct trap modes. The initial Hamiltonian is $\hat H = \hat H^{sp}_0 + \hat H^{int}$ where
\begin{eqnarray}
&&\hat H^{sp}_0=\sum_\alpha \int dx \hat{\psi}^\dagger_\alpha(x)\left(-\frac{1}{2 m_a}\frac{\partial^2}{\partial x^2}+ V^0(x)\right)\hat{\psi}_\alpha(x), \nonumber \\
&&\hat{H}^{int}= \frac{2a_s}{m_a a_\perp^2}\int dx \hat{\rho}_\uparrow(x)\hat{\rho}_\downarrow(x). \nonumber \notag
\end{eqnarray}
$\hat{\psi}_\alpha(x)$ is the fermionic field operator for spin $\alpha$ at point $x$, $a_s$ is the {\it s}-wave scattering length, $\hat\rho_\alpha(x) = \hat{\psi}^\dagger_\alpha(x)\hat{\psi}_\alpha(x)$, $\hbar=1$, and we have integrated over two transverse directions with small confinement length $a_\perp\ll a_H$, with $a_H=(m_a\omega)^{-\frac{1}{2}}$. Note that the initial spin-polarized sample will not experience interactions. A resonant $\pi/2$ pulse collectively rotates the spin to the $X$-axis, and a magnetic field gradient is suddenly turned on. This introduces a sudden change (quench) in the single-particle Hamiltonian $\hat H^{sp}_0$, which becomes spin-dependent, $\hat H^{sp}$, where
 \begin{eqnarray}
 \hat H^{sp}&=&\sum_\alpha \int dx \hat{\psi}^\dagger_\alpha(x)\left(-\frac{1}{2 m_a}\frac{\partial^2}{\partial x^2}+ V^{\alpha}(x)\right)\hat{\psi}_\alpha(x). \notag
\end{eqnarray} 
This quench protocol is illustrated in Fig.~\ref{diagram}(a). The spin-dependence of the trapping potential $V^{\alpha = \uparrow,\downarrow}(x)$ creates an inhomogeneity between the spin species, allowing contact {\it s}-wave collisions to occur. Expanding the field operators in the basis of single-particle eigenstates $\phi_n^{\alpha}(x)$ with associated creation operator $\hat c^{\dagger}_{n\alpha}$ and defining the interaction parameter $u_{\uparrow\downarrow}=2a_s/(m_a a_H a_\perp^2)$, $\hat{H}^{int}$ becomes $u_{\uparrow \downarrow}\sum_{nmpq}A_{nmpq}\hat{c}^\dagger_{n \uparrow}\hat{c}_{m \uparrow}\hat{c}^\dagger_{p \downarrow}\hat{c}_{q \downarrow}$, where ${A_{nmpq} = a_H\int dx \phi_n^\uparrow(x)\phi_m^\uparrow(x)\phi_p^\downarrow(x)\phi_q^\downarrow(x)}$.

To model two classes of experiments \cite{Koschorreck2013, Bardon14, Trotzky15} and \cite{Deutsch2010, Lewandowski2002, Du2008, Du2009}, we consider spin-dependent potentials of the form $V^{\alpha=\uparrow, \downarrow}(x)=V^0(x)+\Delta V^\alpha (x)$, with $ \Delta V^\alpha (x)$ generated by a magnetic field with  a constant gradient, $\Delta V^\alpha(x)=\pm Bx$,  or a linear gradient, $\Delta V^\alpha(x)=\pm m_a\omega_B^2  x^2/2$. In both cases $\hat H^{sp}$ can be written as:
  \begin{eqnarray}\hat H^{sp} &=&  \displaystyle\sum\limits_n \left[\bar\omega(n + 1/2)\hat N_n + \Delta\omega \left(n+1/2\right)\hat \sigma^Z_n \right],\nonumber
\end{eqnarray}  with $\hat N_n = \hat c^{\dagger}_{n \uparrow}\hat c_{n\uparrow}+\hat c^{\dagger}_{n\downarrow}\hat c_{n\downarrow}$,
and $\{\hat\sigma^X_n,\hat\sigma^Y_n,\hat\sigma^Z_n\} \equiv \sum_{\alpha, \beta}\hat{c}^\dagger_{n\alpha}\vec{\sigma}_{\alpha \beta}\hat{c}_{n\beta} $ where  $\vec{\sigma}$ is a vector of  Pauli matrices.
The constant gradient shifts the trap for spin up (down) by $x_0$ ($-x_0$), with $x_0= \frac{B}{m_a\omega^2}$, but does not change the frequency; $\bar\omega=\omega$ and $\Delta\omega=0$. In a noninteracting gas the $\downarrow$ and $\uparrow$ densities and the magnetization oscillate at frequency $\omega$ due to this motion~\cite{Koller2015, xu2015}.  A linear gradient adds an additional harmonic potential term resulting in different trap frequencies for the two spins: $\bar \omega = (\omega^\uparrow +\omega^\downarrow)/2$ and $\Delta \omega = (\omega^\uparrow -\omega^\downarrow)/2$. The non-interacting spin densities undergo a breathing motion in their respective traps, leading to oscillations in the total magnetization~\cite{Koller2015}.  A finite $\Delta \omega$ causes dephasing through rotations of the magnetization in the $XY$ plane with mode-dependent rates.

{\it The generalized spin model approximation--} The quench of the trapping potential to a spin-dependent form projects the initially polarized  state, which we take to be the ground state in this work, onto the eigenmode basis of $\hat H^{sp}$\footnote{The initial $2N$ spin-independent populated modes (${0,...,N-1}$ for both spin-up and spin-down) are projected onto $2\tilde{N}$ modes, where the $\tilde{N}$ modes for spin up are different than the $\tilde{N}$ for spin down, and $\tilde{N}$ is chosen such that the initial state is reproduced to an error of $10^{-16}$ in the norm}. The resulting state $|\psi \rangle _{t=0}$ is a coherent superposition of many product states, each characterized by a set of populated modes ${{\bf n}^i=\{ {\bf n}^i_1, {\bf n}^i_2, \dots, {\bf n}^i_N\}}$:
${|\psi\rangle_{t=0}=\sum_i d_i \prod_{j=1}^{N} \hat{c}^{\dagger}_{\mathbf{n}_j^i \sigma_j}|0\rangle}$ The coefficients $d_i$ are determined by the change of basis associated with the eigenstates of $V^0(x)$ and $V^{\alpha=\uparrow,\downarrow}(x)$.

Our key approximation is that single particle modes either remain the same or are exchanged between two colliding atoms. Exact numerical calculations confirm the validity of this approximation in the weakly interacting regime~\cite{supplement}. For each set ${\bf n}^i$ the resulting total  Hamiltonian  takes the form of an XXZ spin model,
\begin{eqnarray}
\textstyle \hat H^{sm}_{{\bf n}^i}=  \hat H^{sp}_{{\bf n}^i}-\frac{u_{\uparrow\downarrow}}{4}\sum_{n \neq m \in {\bf n}^i}\sum_{\nu=X,Y,Z}J^{\nu}_{nm} \hat{\sigma}^{\nu}_n\hat{\sigma}^{\nu}_m\, ,
\label{smequation}
\end{eqnarray} 
plus additional small density-$\hat \sigma^Z$ couplings~\cite{supplement}.  Here, the Ising, $J^Z_{nm} \equiv A_{nnmm}$, and exchange, $J^{X}_{nm}=J^{Y}_{nm}=J^\perp_{nm} \equiv A_{nmmn}$, couplings result from the overlap between the $\uparrow$ and $\downarrow$ single-particle eigenstates and are long-ranged ($\sim 1/\sqrt{|n-m|}$) in each direction $(x,y,z)$ \cite{supplement}. 
In this approximation, each sector $\mathbf{n}^i$ evolves independently, but with $\mathbf{n}^i$-dependent parameters, under Eq.~\ref{smequation}.  When computing observables, we account for both the interaction-driven spin dynamics within each ${\bf n}^i$ sector, as well as the single particle dynamics determined from the coherences between sectors.

\begin{figure*}
\includegraphics[width=2.0\columnwidth]{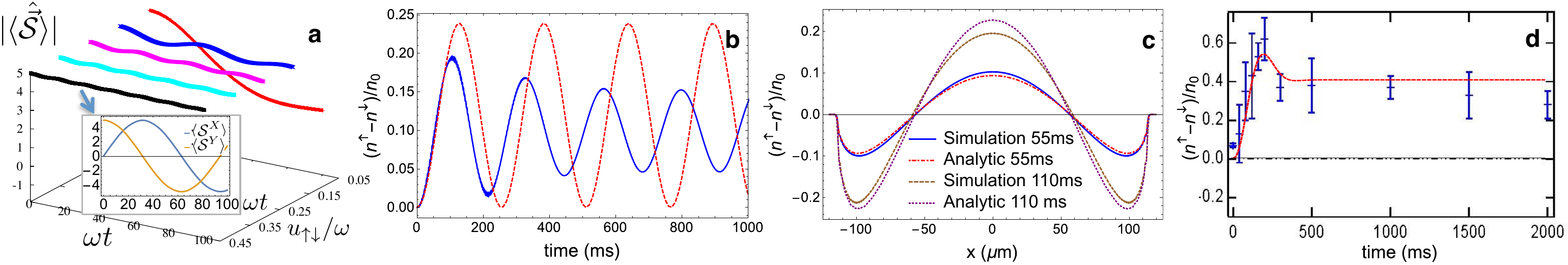} %Retain asterik for wide figure and wide figure captions\includegraphics[scale=.3]{linear_fig.pdf}
\caption
{ \label{linear_fig}
\raggedright
(Color online) Dynamics for a linear gradient. (a) Spin self-rephasing for $\omega_B=0.1\omega$: as interactions increase, demagnetization is suppressed and $\langle \hat{\vec{\mathcal S}}  \rangle$ precesses collectively in the $XY$ plane (inset). (b) Simulation of a one dimensional gas at zero temperature with parameters from Ref.~\cite{Du2008}, showing $(n^\uparrow-n^\downarrow)/n_0$ at the cloud center (blue solid line) with analytic prediction (red dashed line), and (c) segregated spin density profiles. (d) Data from Ref.~\cite{Du2008}, and prediction (red dashed line) based on a thermal average of Rabi oscillations between the Dicke and spin-wave manifolds.}
\end{figure*}

{\it Spin observables--} The local and collective magnetizations are given by $\hat{\vec{\mathcal S}}(x) = \frac{1}{2}\sum_{nm,\alpha,\beta}\phi^\alpha_n(x)\phi^\beta_m(x)\left(\hat{c}^{\dagger \alpha}_n \vec{\sigma}_{\alpha \beta}\hat{c}^\beta_m\right)$ and $\hat{\vec{\mathcal S}} =\int dx \hat{\vec{\mathcal S}}(x)$.  Fig.~\ref{constant_fig} summarizes the results for a constant gradient with $N=10$ \footnote{All simulations displayed in figures in the main text are for $N=10$ except for those in Fig.~\ref{linear_fig}(b-d) which are for $N=560$, $N=560$, and $N=2\times10^5$, respectively.}. At short times the collective magnetization $\langle \hat {\mathcal S}^{X} \rangle$ ((a) and (e)) exhibits characteristic single-particle oscillations at frequency $\omega$; these quickly dephase and are modulated by a global envelope with a longer time scale.  Similar behavior is observed for the local magnetizations $\langle \hat{\mathcal S}^{X,Y,Z} (x) \rangle$ (b-d, f-h).  Although the total $\langle \hat{\mathcal S}^{Y,Z}\rangle$ magnetizations are zero at all times, the local quantities $\langle\hat{\mathcal S}^{Y, Z}(x)\rangle$ evolve due to coherences between mode configurations. Their dynamics, however, are damped by interactions.

The dynamics can be understood as follows.  For spin independent potentials, $J^Z_{nm} = J^\perp_{nm}$ and $\Delta\omega=0$. The  Hamiltonian  $\hat H^{sm}_{{\bf n}^i}$ is SU(2) symmetric and commutes with $\hat{ \vec{S}}^2$, where $\hat{\vec{S}} \equiv \frac{1}{2}\sum_n\hat{\vec{\sigma}}_n$, and so its eigenstates can be labelled by the total spin $S$.  When a gradient is applied, the SU(2) symmetry is broken by terms $\Delta_{nm} = J^Z_{nm}-J^\perp_{nm}$  ($\Delta\omega=0$ for a constant gradient), and the Hamiltonian can be rewritten as $\hat H^{S}_{{\bf n}^i}+\hat H^\delta_{{\bf n}^i}$, where
\begin{eqnarray}
&&\hat H^{S}_{{\bf n}^i}=E_{{\bf n}^i}-\frac{u_{\uparrow \downarrow}}{4}\displaystyle\sum\limits_{n \neq m \in {\bf n}^i}\left[J^\perp_{nm} \vec{\sigma}_n\cdot\vec{\sigma}_m+\bar\Delta\hat\sigma^Z_n \hat\sigma^Z_m\right], \nonumber \\
&&\hat H^\delta_{{\bf n}^i} = -\frac{u_{\uparrow \downarrow}}{4}\displaystyle\sum\limits_{n \neq m \in {\bf n}^i} \delta_{nm} \hat\sigma^Z_n \hat\sigma^Z_m ,
\label{HsHdelta}
\end{eqnarray} $E_{{\bf n}^i}= \bar\omega \sum_{n \in {\bf n}^i} (n + 1/2)$  is a constant, $\bar\Delta$ is the average value of $\Delta_{nm}$, and $\delta_{nm} = \Delta_{nm} - \bar\Delta$. $\hat H^S_{{\bf n}^i}$  commutes with  $\hat {\vec{S}}^2$ so only $\hat H^\delta_{{\bf n}^i}$ induces transitions between manifolds of different $S$. For a sufficiently weak gradient, and   $\delta_{nm} \ll J^\perp_{nm}$, a large energy gap $G$, which we call the \emph{Dicke gap}, opens between the $S=N/2$ ``Dicke" manifold and the $S=(N/2-1)$ ``spin-wave" manifold~\cite{supplement}. The state of the system begins in the Dicke manifold, and it remains there when terms in $\hat H^\delta_{{\bf n}^i}$ are small compared to this gap \cite{Rey2008a}. Dynamics resulting from the collective Ising term in $\hat H^S_{{\bf n}^i}$ is given by $\langle \hat {S}^X \rangle_{{\bf n}^i} =\frac{N}{2}\cos^{N-1}\left(u_{\uparrow \downarrow}\bar{\Delta}t \right),$ and $\langle \hat {S}^{Y,Z} \rangle_{{\bf n}^i} = 0$.  Since the interaction parameters $J^Z_{nm}$ and $J^\perp_{nm}$ vary slowly with parameter index, the dynamics of  $\langle \hat {S}^X \rangle_{{\bf n}^i}$ is approximately the same for all $i$, and a single configuration ${\bf n}^0\equiv \{0,1,\cdots N-1\}$ well reproduces the demagnetization envelope (Fig.~\ref{constant_fig}(a)).

For strong gradients, exchange processes are suppressed and the effective interaction Hamiltonian becomes a generic Ising model $\hat H^{\rm Ising}_{{\bf n}^i} =-\frac{u_{\uparrow\downarrow}}{4}\sum_{n\neq m\in  {{\bf n}^i}}J^Z_{nm} \hat \sigma^Z_n\hat \sigma^Z_m$, which also admits a simple expression for the spin magnetization dynamics \cite{Emch1966, Radin1970, Kastner2011, Foss-Feig2013} $\langle \hat S^X \rangle_{{\bf n}^i}  = \sum_{n \in  {{\bf n}^i}}\prod_{m \neq n \in  {{\bf n}^i}} \cos\left(u_{\uparrow\downarrow}J^Z_{nm} t\right)$. In this limit the demagnetization envelope can be captured by the ${\bf n}^0$ realization of the generic Ising solution (Fig.~\ref{constant_fig}(e)).

Short time dynamics of an XXZ Hamiltonian \cite{Hazzard2014} is given by
${\langle \hat {S}^X \rangle = \langle \hat {S}^X \rangle_{t=0} \left(1-(t/\tau_M)^2\right) + O(t^3)}$,
where we define $\tau_M$ as the demagnetization time. By analyzing the scaling of the interaction parameters we find that
$
\tau_M \sim \left(N u_{\uparrow \downarrow}x_0^2\right)^{-1},$ which agrees well the numerical scaling $\sim u_{\uparrow\downarrow}^{-1}x_0^{-2}N^{-0.823}$~\cite{supplement}. Similar behavior was reported in Ref.~\cite{Koschorreck2013} in the weakly-interacting regime \footnote{We note that the spin echo pulse applied  in Refs.~\cite{Koschorreck2013,Bardon14} modifies the single-particle physics~\cite{Koller2015}, but does not affect the interaction-induced collective demagnetization}. 
\begin{figure} %Retain asterik for wide figure and wide figure captions
\includegraphics[width=1.0\columnwidth]{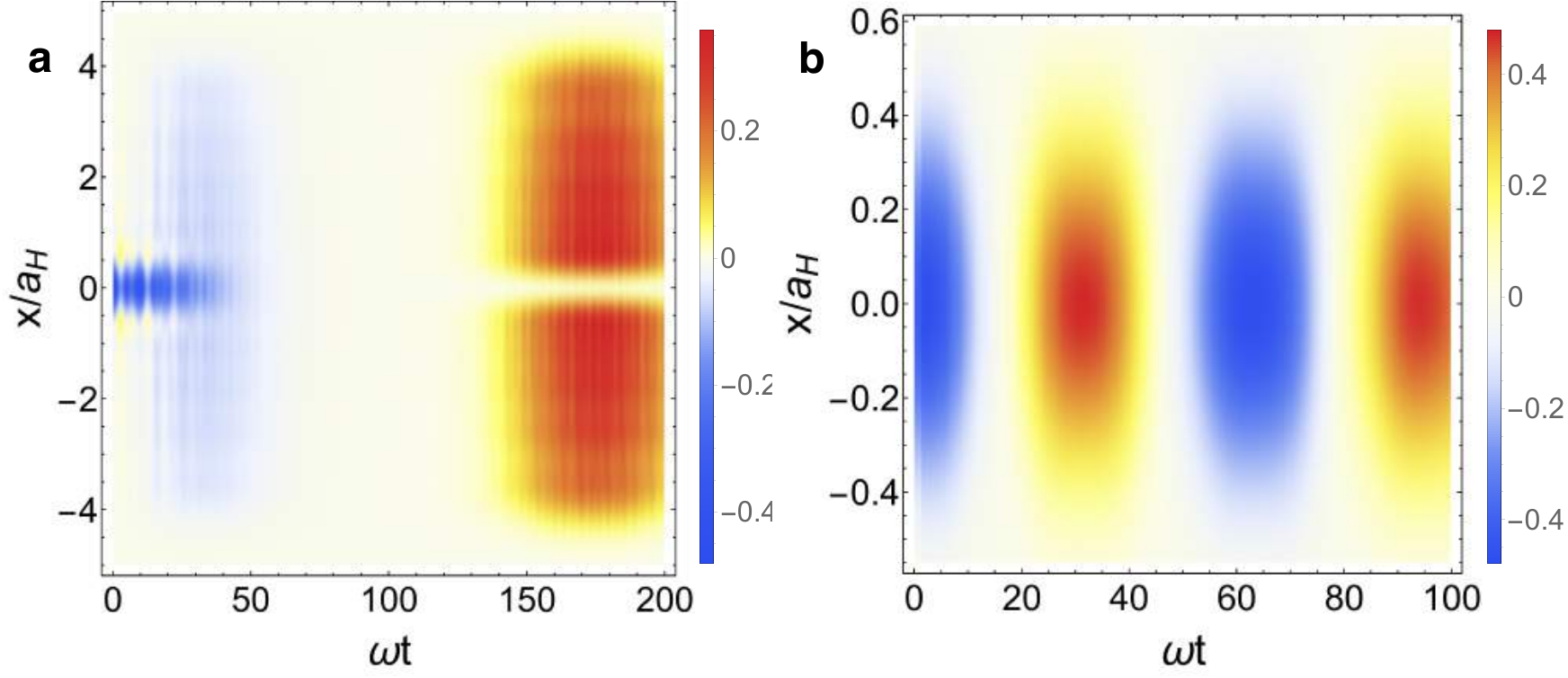}
{\caption
{\raggedright\label{correlations} (Color online) (a) Real part of the connected correlation function ${\rm Re}\left[G^{++}(x,0; t)\right]$ for a weak gradient ($x_0=0.1a_H, u_{\uparrow,\downarrow}=0.35\omega$). Correlations grow collectively due to the long-ranged nature of the interactions in energy space, and peak when the gas is demagnetized. (b) For a linear gradient in the self-rephasing regime ($\omega_B=0.1\omega, u_{\uparrow\downarrow}=0.45\omega$), the connected correlator ${\rm Re}\left[G^{++}(x,0; t)\right]$ rotates collectively in the $XY$ plane.}}
\end{figure}

Fig.~\ref{linear_fig} (a) shows the numerically-obtained total magnetization vs. interactions for a weak linear gradient. The magnetization remains nearly constant for sufficiently strong interactions, and the collective spin dynamics is a global precession in the $XY$ plane (inset). This self-rephasing effect was experimentally reported in Ref.~\cite{Deutsch2010}, and the spin model provides a simple interpretation. For a system in a weak gradient, the single-particle term $\propto\Delta \omega$ is the largest inhomogeneity. In this limit the Hamiltonian simplifies to ${-\frac{u_{\uparrow \downarrow}}{4}\sum_{n \neq m}J^\perp_{nm} \vec{\sigma}_n\cdot\vec{\sigma}_m +  \sum_n \Delta\omega (n+\frac{1}{2})\hat \sigma^Z_n}$. When $\Delta\omega N_{{\bf n}^i}^{{\rm ave}} \ll G$, where $G$ is the Dicke gap and $N_{{\bf n}^i}^{{\rm ave}}$ is the average mode occupation, most of the population remains in the Dicke manifold. After projecting $\hat H^{sp}$ onto the Dicke states, the dynamics is a collective precession in the $XY$ plane of the generalized Bloch vector, i.e $\langle {\hat{S}}^\pm(t)  \rangle= \langle {\hat{{S}}}^\pm(0) \rangle e^{\pm 2 it(N_{{\bf n}^i}^{{\rm ave}} + \frac{1}{2})  \Delta\omega},$ with ${ \hat{S}}^\pm={\hat{S}}^X\pm  i{\hat{{S}}}^Y$. Demagnetization is suppressed when interactions ($\propto G$) dominate over the dephasing introduced by $\Delta\omega$. Under this condition, a large fraction of the population stays in the Dicke manifold.

Spin segregation in fermionic gases -- a clear, spatial separation of the spin densities, first reported in Ref.~\cite{Du2008} -- occurs at timescales set by the mean interaction energy, and reverses sign when interactions are switched from attractive to repulsive. When $\Delta\omega N \ll G$, this effect can be understood as the result of off-resonant Rabi oscillations between the $S=N/2$ Dicke states and the $S=(N/2-1)$ spin-wave states, which are coupled by the gradient and whose energies are separated by the Dicke gap $G$. If the gradient is weak, one can ignore coherences developed between mode sectors, and approximate $\phi^\uparrow_n(x)\approx\phi^\downarrow_n(x)=\phi_n(x)$. In this limit the dynamics of the population difference $\Delta n = n^\uparrow(x)-n^\downarrow(x)$ is approximately \cite{supplement}
\begin{eqnarray}
\langle \Delta n \rangle= \frac{2\Delta\omega}{G}\sum_{n\in{\bf n}^i}\phi_n(x)^2\left(n-N_{{\bf n}^i}^{{\rm ave}}\right)\left(\cos\left(Gt\right)-1\right). \label{sigmazn1}
\end{eqnarray}
The spin density changes sign when $n>N_{{\bf n}^i}^{{\rm ave}}$. Spin segregation occurs as a result since high energy modes on average occupy positions further from the origin than low energy modes.

We now proceed to use the spin model framework  to model the segregation observed in Ref.~\cite{Du2008}. Although the measurements were done in the high temperature regime, we first determine the role of single particle motion by modeling a simpler 1D case at zero temperature with the same effective parameters. This case can be exactly solved with t-DMRG \cite{supplement} and Figs.~\ref{linear_fig}(b,c) show the dynamics of $(n^\uparrow(x)-n^\downarrow(x))/n_0$, where $n_0=(n^\uparrow(0)+n^\downarrow(0))/2$. Single particle motion is negligible, and the dynamics is closely approximated by Eq.~\ref{sigmazn1}. This information allows us to model the actual experiment with a pure spin model. At the high temperature of the experiment, the Dicke gap significantly decreases, however, Eqn.~\ref{sigmazn1} remains valid at short times when the majority of the population is in the Dicke manifold. The segregation obtained from a thermal average of Eqn.~\ref{sigmazn1} \cite{supplement} well reproduces the experiment as shown in Fig. 3d. For this calculation the only free parameter is the asymptotic value of the density imbalance \footnote{The asymptotic value of the spin density imbalance is chosen to be 0.4, which matches the experimental values from 500-1000ms. Relaxation due to other decoherence mechanisms occurs at $\sim$2s.}. The population difference saturates due to dephasing associated with the thermal spread of the $G$ values.

{\it Correlations--} Our approach can be used to compute higher-order correlations, such as the  $G^{++}(x,x') =\langle \hat{ \mathcal {S}}^+(x)\hat{ \mathcal {S}}^+(x') \rangle - \langle \hat {\mathcal{S}}^+(x) \rangle \langle \hat{ \mathcal{ S}}^+(x')\rangle$ correlator shown in  Fig.~\ref{correlations}.  Although the system is initially non-interacting, $G^{++}(t=0)$ shows finite anti-bunching correlations near $x\sim x'$ arising from Fermi statistics (mode entanglement) \cite{vedral2003, clark2005}.  At later times, correlations behave collectively, a distinct consequence of the long-range character of the spin coupling parameters \cite{hauke2013, schachenmayer2013, eisert2013, gong2014, richerme2014}.

For a weak constant gradient, the collective Ising model provides a good characterization of the correlation dynamics. For each spin configuration $G_{{\bf n}^i}^{++}(x,x';t)= {f_1^i(x,x')\cos^{N-2}\left(2u_{\uparrow \downarrow}\bar\Delta t\right)  -f_2^i(x,x')\cos^{2N-2}\left(u_{\uparrow \downarrow}\bar\Delta t\right)}$, where the functions $f_{1,2}^i(x,x')$ depend on the set of populated modes \cite{supplement}. $G^{++}$ peaks at the time when the system has completely demagnetized (Fig.~\ref{correlations}(a)). For a pure spin system with a collective Ising Hamiltonian, the state at this time is a Schr{\"o}dinger-cat state \cite{kitagawa1993, opatrny2012}. For the linear gradient in the self-rephasing regime, we observe collective precession of $G^{++}$ (Fig.~\ref{correlations}(b)).
As interactions decrease or the inhomogeneity increases, correlations are strongly affected by the interplay between single-particle dynamics and interactions. Mode entanglement tends to cause an almost linear spreading of the correlations with time~\cite{Lieb1972, nachtergaele2006, cheneau2012}, while interactions tend to globally distribute  and damp those  correlations \cite{supplement}. Current experiments are in position to confirm these predictions.

{\it Outlook--} We have discussed an approach to model the interplay of motional and spin degrees of freedom in weakly interacting fermionic systems in spin-dependent potentials. Simulations reproduce several collective dynamical phenomena that were recently observed in cold gas experiments, and we can understand the physics behind these effects with simple considerations. For larger systems and in higher dimensions, methods such as the discrete truncated Wigner approximation could be utilized \cite{Polkovnikov2010, Schachenmayer2015, Pucci2015,1367-2630-17-6-065009}.  Our formulation may also be useful for modeling other spin transport experiments~\cite{Sommer2011, Niroomand2015}.
\section{Acknowledgements}
We thank J. E. Thomas, K. R. A. Hazzard, A. Pikovski, and J. Schachenmayer for useful discussions, and P. Romatschke, J. Bohnet, and M. G{\"a}rttner for comments on the manuscript.
This work was supported by JILA-NSF-PFC-1125844, NSF-PIF-
1211914, ARO, AFOSR, and AFOSR-MURI. AK was supported by the Department of Defense through the NDSEG program.  MLW thanks the NRC postdoctoral fellowship program for support.

\bibliography{spinsegcitations7_withAppendix2}

\begin{thebibliography}{10}

\bibitem{CMR}
A.~P. Ramirez, Journal of Physics: Condensed Matter {\bf 9},  8171  (1997).

\bibitem{Koschorreck2013}
M. Koschorreck, D. Pertot, E. Vogt, and M. Kohl, Nature Physics {\bf 9},  405
  (2013).

\bibitem{Bardon14}
A.~B. Bardon, S. Beattie, C. Luciuk, W. Cairncross, D. Fine, N.~S. Cheng,
  G.~J.~A. Edge, E. Taylor, S. Zhang, S. Trotzky, and J.~H. Thywissen, Science
  {\bf 344},  722  (2014).

\bibitem{Trotzky15}
S. Trotzky, S. Beattie, C. Luciuk, S. Smale, B. Bardon, A.\, T. Enss, E.
  Taylor, S. Zhang, and H. Thywissen, J.\, Phys. Rev. Lett. {\bf 114},  015301
  (2015).

\bibitem{Deutsch2010}
C. Deutsch, F. Ramirez-Martinez, C. Lacro\^ute, F. Reinhard, T. Schneider,
  J.~N. Fuchs, F. Pi\'echon, F. Lalo\"e, J. Reichel, and P. Rosenbusch, Phys.
  Rev. Lett. {\bf 105},  020401  (2010).

\bibitem{Lewandowski2002}
H.~J. Lewandowski, D.~M. Harber, D.~L. Whitaker, and E.~A. Cornell, Phys. Rev.
  Lett. {\bf 88},  070403  (2002).

\bibitem{Du2008}
X. Du, L. Luo, B. Clancy, and J.~E. Thomas, Phys. Rev. Lett. {\bf 101},  150401
   (2008).

\bibitem{Du2009}
X. Du, Y. Zhang, J. Petricka, and J.~E. Thomas, Phys. Rev. Lett. {\bf 103},
  010401  (2009).

\bibitem{Fuchs2002}
J.~N. Fuchs, D.~M. Gangardt, and F. Lalo\"e, Phys. Rev. Lett. {\bf 88},  230404
   (2002).

\bibitem{Williams2002}
J.~E. Williams, T. Nikuni, and C.~W. Clark, Phys. Rev. Lett. {\bf 88},  230405
  (2002).

\bibitem{Bradley2002}
A.~S. Bradley and C.~W. Gardiner, Journal of Physics B: Atomic, Molecular and
  Optical Physics {\bf 35},  4299  (2002).

\bibitem{Natu2009}
S.~S. Natu and E.~J. Mueller, Phys. Rev. A {\bf 79},  051601  (2009).

\bibitem{ebling11}
U. Ebling, A. Eckardt, and M. Lewenstein, Phys. Rev. A {\bf 84},  063607
  (2011).

\bibitem{Bruun2011}
G.~M. Bruun, New Journal of Physics {\bf 13},  035005  (2011).

\bibitem{Piechon2009}
F. Pi\'echon, J.~N. Fuchs, and F. Lalo\"e, Phys. Rev. Lett. {\bf 102},  215301
  (2009).

\bibitem{xu2015}
J. Xu, Q. Gu, and E.~J. Mueller, Phys. Rev. A {\bf 91},  043613  (2015).

\bibitem{Goulko2013}
O. Goulko, F. Chevy, and C. Lobo, Phys. Rev. Lett. {\bf 111},  190402  (2013).

\bibitem{Enss2015}
T. Enss, Phys. Rev. A {\bf 91},  023614  (2015).

\bibitem{Oktel2002}
M.~O. Oktel and L.~S. Levitov, Phys. Rev. Lett. {\bf 88},  230403  (2002).

\bibitem{Gibble2009}
K. Gibble, Physical Review Letters {\bf 103},  113202  (2009).

\bibitem{Rey2009}
{ A. M. Rey }and A.~V.~Gorshkov and C. Rubbo, Phys. Rev. Lett. {\bf 103},
  260402  (2009).

\bibitem{Yu2010}
Z.~H. Yu and C.~J. Pethick, Phys. Rev. Lett. {\bf 104},  010801  (2010).

\bibitem{Hazlett2013}
E. Hazlett, Y. Zhang, R. Stites, K. Gibble, and K.~M. O'Hara, Phys. Rev. Lett.
  {\bf 110},  160801  (2013).

\bibitem{Koller14}
A.~P. Koller, M. Beverland, A.~V. Gorshkov, and A.~M. Rey, Phys. Rev. Lett.
  {\bf 112},  123001  (2014).

\bibitem{Beverland14}
M.~E. Beverland, G. Alagic, M.~J. Martin, A.~P. Koller, A.~M. Rey, and A.~V.
  Gorshkov, arXiv:1409.3234  (20014).

\bibitem{Swallows2011}
M.~D. Swallows, M. Bishof, Y.~G. Lin, S. Blatt, M.~J. Martin, A.~M. Rey, and J.
  Ye, Science {\bf 331},  1043  (2011).

\bibitem{Maineult2012}
W. Maineult, C. Deutsch, K. Gibble, J. Reichel, and P. Rosenbusch, Phys. Rev.
  Lett. {\bf 109},  020407  (2012).

\bibitem{Martin2013}
M.~J. Martin, M. Bishof, M.~D. Swallows, X. Zhang, C. Benko, J. von Stecher,
  A.~V. Gorshkov, A.~M. Rey, and J. Ye, Science {\bf 341},  632  (2013).

\bibitem{Pechkis2013}
H. Pechkis, J. Wrubel, A. Schwettmann, P. Griffin, R. Barnett, E. Tiesinga, and
  P. Lett, Phys. Rev. Lett. {\bf 111},  025301  (2013).

\bibitem{Yan2013}
B. Yan, S.~A. Moses, B. Gadway, J.~P. Covey, K.~R.~A. Hazzard, A.~M. Rey, D.~S.
  Jin, and J. Ye, Nature {\bf 501},  521  (2013).

\bibitem{Sommer2011}
A. Sommer, M. Ku, G. Roati, and M.~W. Zwierlein, Nature {\bf 472},  201
  (2011).

\bibitem{Makotyn2014}
P. Makotyn, C.~E. Klauss, D.~L. Goldberger, E. Cornell, and D.~S. Jin, Nature
  Physics {\bf 10},  116  (2014).

\bibitem{Schollwoeck}
U. Schollw{\"o}ck, Annals of Physics {\bf 326},  96   (2011), january 2011
  Special Issue.

\bibitem{Polkovnikov2010}
A. Polkovnikov, Annals of Physics {\bf 325},  1790  (2010).

\bibitem{Schachenmayer2015}
J. Schachenmayer, A. Pikovski, and A.~M. Rey, Phys. Rev. X {\bf 5},  011022
  (2015).

\bibitem{Pucci2015}
L. Pucci, A. Roy, and M. Kastner, arXiv:1510.03768  (20015).

\bibitem{Emch1966}
G.~G. Emch, Journal of Mathematical Physics {\bf 7},    (1966).

\bibitem{Radin1970}
C. Radin, Journal of Mathematical Physics {\bf 11},    (1970).

\bibitem{Kastner2011}
M. Kastner, Phys. Rev. Lett. {\bf 106},  130601  (2011).

\bibitem{Foss-Feig2013}
M. Foss-Feig, K.~R.~A. Hazzard, J.~J. Bollinger, and A.~M. Rey, Phys. Rev. A
  {\bf 87},  042101  (2013).

\bibitem{Note1}
The matrix product state studies of the main text were performed using
  extensions of the open source MPS library~\cite {OSMPS,Wall_Carr_12}, and are
  described further in the supplement~\cite {supplement}.

\bibitem{Koller2015}
A.~P. Koller, J. Mundinger, M.~L. Wall, and A.~M. Rey, Phys. Rev. A {\bf 92},
  033608  (2015).

\bibitem{Note2}
The initial $2N$ spin-independent populated modes (${0,...,N-1}$ for both
  spin-up and spin-down) are projected onto $2\protect \mathaccentV
  {tilde}07E{N}$ modes, where the $\protect \mathaccentV {tilde}07E{N}$ modes
  for spin up are different than the $\protect \mathaccentV {tilde}07E{N}$ for
  spin down, and $\protect \mathaccentV {tilde}07E{N}$ is chosen such that the
  initial state is reproduced to an error of $10^{-16}$ in the norm.

\bibitem{supplement}
A.~P. Koller, M.~L. Wall, J. Mundinger, and A.~M. Rey, Supplemental material
  (2015).

\bibitem{Note3}
All simulations displayed in figures in the main text are for $N=10$ except for
  those in Fig.~\ref {linear_fig}(b-d) which are for $N=560$, $N=560$, and
  $N=2\times 10^5$, respectively.

\bibitem{Rey2008a}
A.~M. Rey, L. Jiang, M. Fleischhauer, E. Demler, and M.~D. Lukin, Phys. Rev. A
  {\bf 77},  052305  (2008).

\bibitem{Hazzard2014}
K.~R.~A. Hazzard, M. van~den Worm, M. Foss-Feig, S.~R. Manmana, E.~G.
  Dalla~Torre, T. Pfau, M. Kastner, and A.~M. Rey, Phys. Rev. A {\bf 90},
  063622  (2014).

\bibitem{Note4}
We note that the spin echo pulse applied in Refs.~\cite
  {Koschorreck2013,Bardon14} modifies the single-particle physics~\cite
  {Koller2015}, but does not affect the interaction-induced collective
  demagnetization.

\bibitem{Note5}
The asymptotic value of the spin density imbalance is chosen to be 0.4, which
  matches the experimental values from 500-1000ms. Relaxation due to other
  decoherence mechanisms occurs at $\sim $2s.

\bibitem{vedral2003}
V. Vedral, Open Physics {\bf 1},  289  (2003).

\bibitem{clark2005}
S. Clark, C.~M. Alves, and D. Jaksch, New Journal of Physics {\bf 7},  124
  (2005).

\bibitem{hauke2013}
P. Hauke and L. Tagliacozzo, Physical review letters {\bf 111},  207202
  (2013).

\bibitem{schachenmayer2013}
J. Schachenmayer, B. Lanyon, C. Roos, and A. Daley, Physical Review X {\bf 3},
  031015  (2013).

\bibitem{eisert2013}
J. Eisert, M. van~den Worm, S.~R. Manmana, and M. Kastner, Physical review
  letters {\bf 111},  260401  (2013).

\bibitem{gong2014}
Z.-X. Gong, M. Foss-Feig, S. Michalakis, and A.~V. Gorshkov, Physical review
  letters {\bf 113},  030602  (2014).

\bibitem{richerme2014}
P. Richerme, Z.-X. Gong, A. Lee, C. Senko, J. Smith, M. Foss-Feig, S.
  Michalakis, A.~V. Gorshkov, and C. Monroe, Nature {\bf 511},  198  (2014).

\bibitem{kitagawa1993}
M. Kitagawa and M. Ueda, Physical Review A {\bf 47},  5138  (1993).

\bibitem{opatrny2012}
T. Opatrn{\`y} and K. M{\o}lmer, Physical Review A {\bf 86},  023845  (2012).

\bibitem{Lieb1972}
E. Lieb and R. D., Commun. Math. Phys. {\bf 28},  251  (1972).

\bibitem{nachtergaele2006}
B. Nachtergaele, Y. Ogata, and R. Sims, Journal of statistical physics {\bf
  124},  1  (2006).

\bibitem{cheneau2012}
M. Cheneau, P. Barmettler, D. Poletti, M. Endres, P. Schau{\ss}, T. Fukuhara,
  C. Gross, I. Bloch, C. Kollath, and S. Kuhr, Nature {\bf 481},  484  (2012).

\bibitem{1367-2630-17-6-065009}
J. Schachenmayer, A. Pikovski, and A.~M. Rey, New Journal of Physics {\bf 17},
  065009  (2015).

\bibitem{Niroomand2015}
D. Niroomand, S.~D. Graham, and J.~M. McGuirk, Phys. Rev. Lett. {\bf 115},
  075302  (2015).

\bibitem{OSMPS}
Open Source MPS, \href{http://sourceforge.net/projects/openmps/}{
  http://sourceforge.net/projects/openmps/}.

\bibitem{Wall_Carr_12}
M.~L. Wall and L.~D. Carr, New Journal of Physics {\bf 14},  125015  (2012).

\bibitem{Krauser2013}
J.~S. Krauser, U. Ebling, N. FlŠschner, J. Heinze, K. Sengstock, M. Lewenstein,
  A. Eckardt, and C. Becker, Science {\bf 343},  157  (2014).

\bibitem{Zaletel}
M.~P. Zaletel, R.~S.~K. Mong, C. Karrasch, J.~E. Moore, and F. Pollmann, Phys.
  Rev. B {\bf 91},  165112  (2015).

\bibitem{Crosswhite}
G.~M. Crosswhite, A.~C. Doherty, and G. Vidal, Phys. Rev. B {\bf 78},  035116
  (2008).

\bibitem{Murg}
B. Pirvu, V. Murg, J.~I. Cirac, and F. Verstraete, New Journal of Physics {\bf
  12},  025012  (2010).

\end{thebibliography}
\onecolumngrid

\renewcommand{\theequation}{S\arabic{equation}}
\setcounter{equation}{0}
\hrulefill
\begin{center}
\Large\textbf{Supplemental material for ``Dynamics of interacting fermions in spin-dependent potentials''}
\end{center}
\renewcommand \thesection {\Roman{section}}
\renewcommand \thesubsection {\alph{subsection}}
\setcounter{section}{0}
In this supplemental material we discuss the generalized spin model approximation and its range of validity, explain in detail how spin segregation arises in a many body system, give details of our comparison with Ref.~\cite{Du2008}, present dynamical scaling results, and discuss our numerical methods.

\section{ The generalized spin model approximation: validity and discussion}
The spin model approximation ignores interaction-induced changes of the single-particle motional quantum states and is thus only valid when interactions are weak compared to the harmonic oscillator energy spacing, $u_{\uparrow \downarrow} \ll \omega$. The range of validity of this approximation is essentially when the system is ``collisionless," although the exact crossover to the collisional regime depends not only on the interaction energy  but also  on the strength of the gradient for  the quenches  discussed in this work~\cite{xu2015}.  When interactions are weak compared to the oscillator spacing, collisional processes that do not conserve single particle energy can safely be ignored. However, processes that {\it do} conserve single particle energy, but at the same time change the populated single particle modes, i.e. ``resonant" mode changes, can be important for a harmonic trap  \cite{Koller14}. While there are a large number of such terms in a harmonic trap due to the equal spacing of energy levels, realistic optical traps in cold atom experiments include anharmonicity which breaks these degeneracies. In higher dimensions, the non-separability of the trapping potential suppresses the redistribution of energy modes in the transverse directions. When the energy differences due to anharmonicity and non-separability of the trapping potential are larger than the interaction strength, these terms will be suppressed. This was shown to be the case for example in Refs.~\cite{Hazlett2013,Martin2013,Pechkis2013} where a pure spin model accurately described the experimental observations.  Additionally, at very low temperatures, Pauli blocking can partially prevent mode changing collisions for a spin-polarized sample, as recently observed in Ref.~\cite{Krauser2013}. However, even in a spin-polarized gas, spin- and mode-changing processes may occur, resulting in a doubly occupied mode. 

\begin{figure} %Retain asterik for wide figure and wide figure captions
\includegraphics[scale=.6]{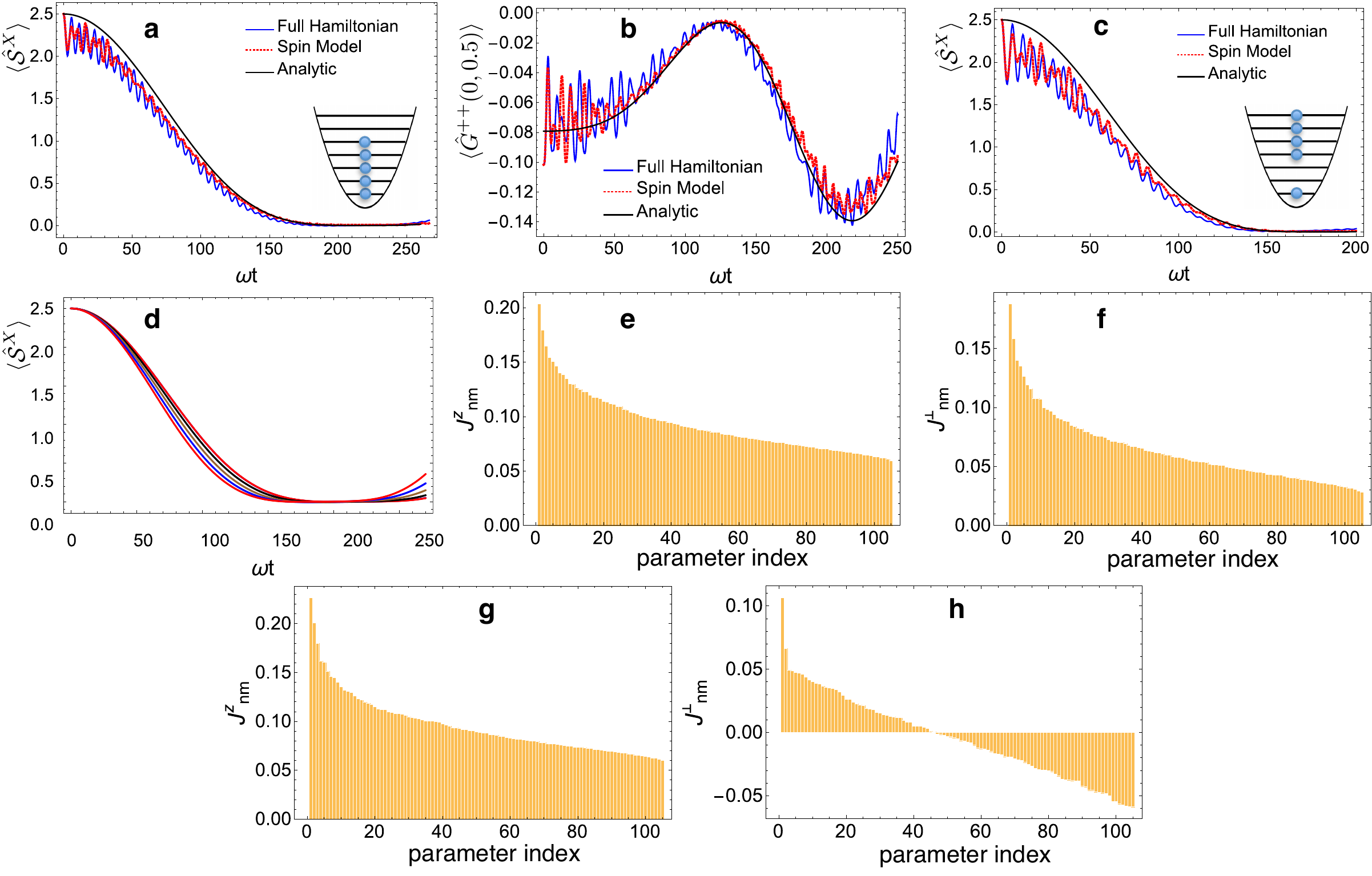}
\caption
{ \label{testspinmodel}
\raggedright
 Spin model approximation vs. full Hamiltonian for 5 particles with $x_0=0.1a$ and $u_{\uparrow \downarrow} = 0.35\omega$. (a) $\langle \mathcal{\hat S}^X \rangle$ quench dynamics for initial modes \{0,1,2,3,4\}, representing a zero temperature gas, along with (b) the connected correlator $G^{++}(x=0, x'=0.5a_H)$. Single-particle oscillations are damped by interactions, and the long time dynamics is well-reproduced by the spin model approximation with decay envelope given by the collective Ising solutions. (c) Dynamics for initial modes \{0,3,4,5,6\} representing a more dilute gas. (d) Dynamics of a pure XXZ spin Hamiltonian with the same parameters, for each of the lowest ``one-hole" mode configurations. The dynamics of each configuration is very similar, explaining why the dynamics of a quench -- involving many configurations -- can be approximated by a single configuration. The interaction parameters vary slowly with parameter index, as shown in (e,f) for $x_0=0.1a_H$ and (g,h) for $x_0 = 0.3a_H$.
}
\end{figure}

We compare exact diagonalization of the full Hamiltonian, including all interaction-induced mode changes, to the spin model prediction for a small number of particles to test its validity. The results are shown in Fig.~\ref{testspinmodel}. Panel (a) shows the dynamics of $\langle \mathcal{\hat S}^X \rangle$ for five particles following a quench of a constant gradient with $x_0=0.1a$ and $u_{\uparrow \downarrow} = 0.35\omega$. The quench induces single-particle dynamics which we observe as fast oscillations at the trapping period. In the spin model approximation, these oscillations are modified due to interactions and become damped at long times. The long time demagnetization and damping of single particle oscillations are well captured by the spin model approximation. Also plotted is the analytic solution for the collective Ising model which captures the demagnetization envelope. Fig.~\ref{testspinmodel}(c) shows the dynamics for a different initial mode configuration -- $\{0,3,4,5,6\}$ -- where Pauli blocking would not prevent several resonant mode changing processes. For instance, the process $(n=0, m=3) \rightarrow (n=1, m=2)$ is resonant. The spin model approximation works well even in this case.

The initial state after a quench is a superposition of many different product states of spins, in different mode configurations labeled ${\bf n}^i$. Because the interaction parameters vary slowly with parameter index, each ${\bf n}^i$ has similar interaction parameters and similar dynamics. Fig.~\ref{testspinmodel}(d) shows the dynamics for 5 spins evolved under a pure XXZ Hamiltonian, with the same conditions as the dynamics in Fig.~\ref{testspinmodel}(a). Each curve represents a different ``one-hole" mode configuration of five spins that differs from ${\bf n}^0 \equiv \{0,1,2,3,4\}$ by exactly one mode (${\bf n}^0$ dynamics is also shown). For instance, the initially occupied modes are $\{0,1,2,3,5\}$ or $\{0,1,2,4,5\}$, etc. All these configurations contribute to the dynamics after a quench. Since they all have similar dynamics, however, we only need to consider the ${\bf n}^0$ configuration to reproduce the demagnetization envelope. The slow variation of the interaction parameters is illustrated in Fig.~\ref{testspinmodel}(e,f) where we plot the value of all the parameters $J^Z_{nm}$ and $J^{\perp}_{nm}$ for modes $n,m=0$ through $n,m=15$, sorted by value and labeled by a parameter index. In Fig.~\ref{testspinmodel}(g,h) we show that the interaction parameters also vary slowly for a stronger gradient, $x_0=0.3a_H$. The slow variation of interaction parameters also helps explain why mode changes are relatively unimportant: a mode change simply evolves the system to another mode configuration where the dynamics are nearly the same.

The collective Ising solution gives the connected correlation function studied in the main text as ${G_{{\bf n}^i}^{++}(x,x';t)= f_1^i(x,x')\cos^{N-2}\left(2u_{\uparrow \downarrow}\bar\Delta t\right)  -f_2^i(x,x')\cos^{2N-2}\left(u_{\uparrow \downarrow}\bar\Delta t\right)}$, where the functions $f_{1,2}^i(x,x')$ are given by
\begin{eqnarray}
&&f_1^i(x,x')=\frac{1}{4}\displaystyle\sum\limits_{nm\in\mathbf{n}^i} \left(\phi^\uparrow_n(x)\phi^\downarrow_n(x)\phi^\uparrow_m(x')\phi^\downarrow_m(x')-\phi^\uparrow_n(x)\phi^\downarrow_n(x')\phi^\uparrow_m(x')\phi^\downarrow_m(x)\right), \nonumber \\
&&f_2^i(x,x')=\frac{1}{4}\displaystyle\sum\limits_{nm\in\mathbf{n}^i}\phi^\uparrow_n(x)\phi^\downarrow_n(x)\phi^\uparrow_m(x')\phi^\downarrow_m(x'). \label{gplusplus}
\end{eqnarray} 
In Fig.~\ref{testspinmodel}(b) we show the connected correlator $G^{++}(x,x')$ evaluated at $x=0, x'=0.5a_H$, along with the analytic solution for the ${\bf n}^0$ mode configuration. The spin model approximation and analytic solution do an excellent job of reproducing the dynamics of the correlation function.
For stronger gradients where the generic Ising model is a better description of the dynamics, 
\begin{align}
\nonumber G_{{\bf n}^i}^{++}(x,x';t)&= \frac{1}{4}\sum_{n,m\in {{\bf n}^i}}\left(\phi^\uparrow_n(x)\phi^\downarrow_n(x)\phi^\uparrow_m(x')\phi^\downarrow_m(x')-\phi^\uparrow_n(x)\phi^\downarrow_n(x')\phi^\uparrow_m(x')\phi^\downarrow_m(x)\right)\prod_{p \neq n,m\in\mathbf{n}^i}\cos\left(J^Z_{np}t + J^Z_{mp}t\right)\\
&-\frac{1}{4}\left[\sum_{n\in {{\bf n}^i}}  \left(\phi^\uparrow_n(x)\phi^\downarrow_n(x)\right)\prod_{p\ne n\in \mathbf{n}^i}\cos\left(J^Z_{np}t\right)\right]\left[\sum_{n\in {{\bf n}^i}}  \left(\phi^\uparrow_n(x')\phi^\downarrow_n(x')\right)\prod_{p\ne n\in \mathbf{n}^i}\cos\left(J^Z_{np}t\right)\right]\, .
\end{align}
\begin{figure*} %Retain asterik for wide figure and wide figure captions
\begin{centering}
\includegraphics[scale=0.45]{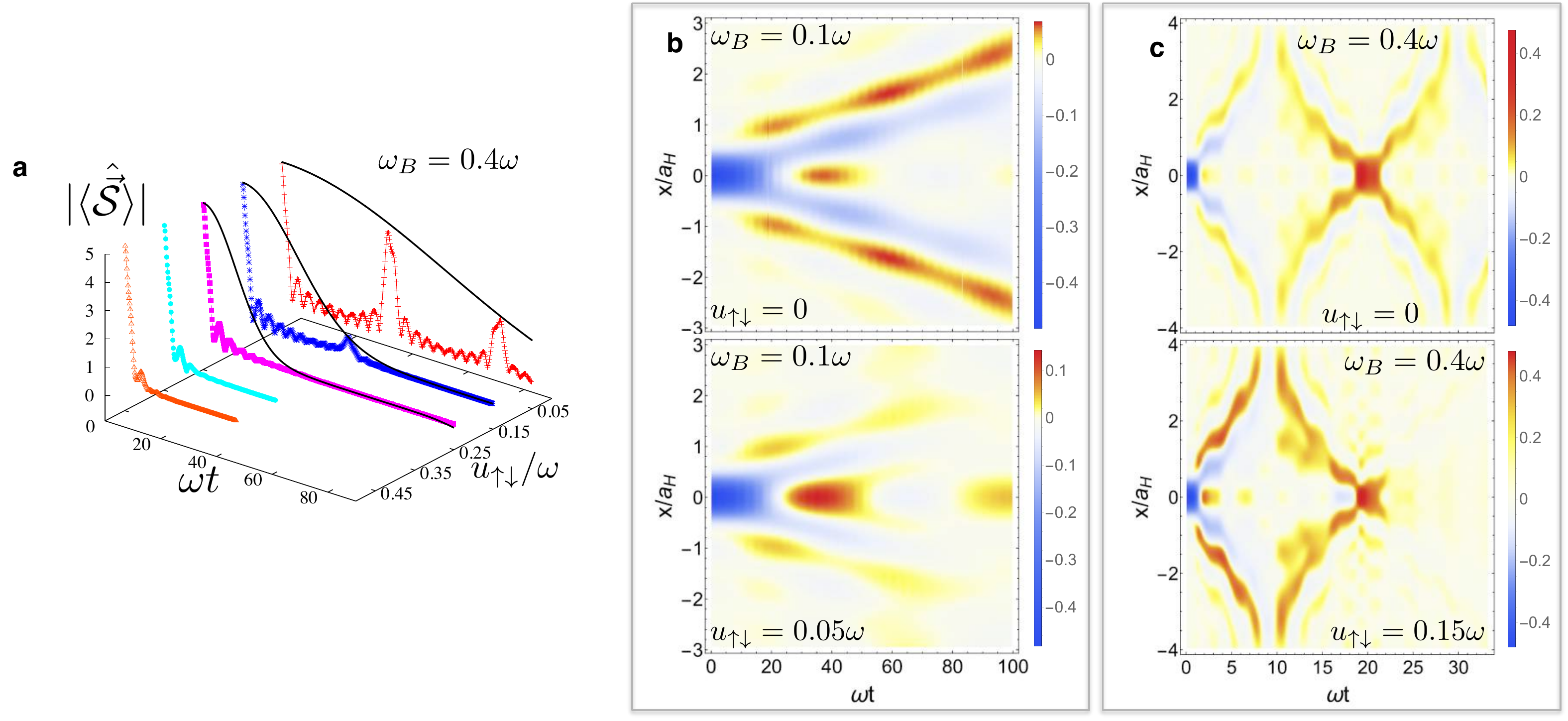}
\caption
{ \label{extra_linear}
\raggedright
(a) Demagnetization dynamics for a strong linear gradient. The envelopes are given by the generic Ising solutions (black lines). (b)  For weak interactions or strong gradients (c, d), interactions collectively damp the correlations arising from quantum statistics (upper panels are non-interacting).}
\end{centering}
\end{figure*}

As discussed in the main text, for strong linear gradients exchange is suppressed and the demagnetization envelope is given by the generic Ising solutions. These simulations and predictions are shown in Fig.~\ref{extra_linear}(a). For the linear gradient in the self-rephasing regime, we observe collective precession of the correlation function $G^{++}$, seen in Fig.~4(b) in the main text.
As interactions decrease or the inhomogeneity increases, mode entanglement tends to cause an almost linear spreading of the correlations with time~\cite{Lieb1972, nachtergaele2006, cheneau2012}. Interactions tend to globally distribute and damp those correlations, as seen in Fig.~\ref{extra_linear}(c, d).

\begin{figure*} %Retain asterik for wide figure and wide figure captions
\begin{centering}
\includegraphics[scale=0.6]{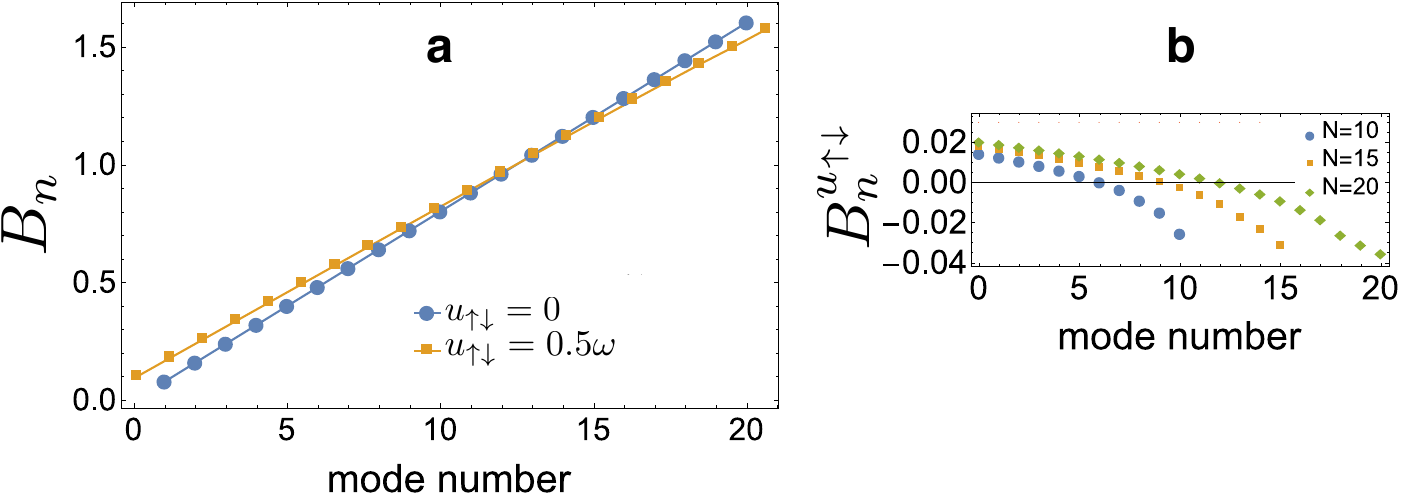}
\caption
{ \label{zfield}
\raggedright
(a) Magnitude of of the total field $B_n\hat\sigma^Z_n $, which contains both single particle ($B^{sp}\hat\sigma^Z_n$) and interaction ($B_n^{u_{\uparrow \downarrow}}\hat\sigma^Z_n$) terms, for a linear gradient with $\Delta\omega = 0.08\omega$. Even for strong interactions ($u_{\uparrow \downarrow}=0.5\omega$), the Hamiltonian is not significantly modified by the interaction-induced terms $B_n^{u_{\uparrow \downarrow}}$ which appear when $J^Z_{nm} \neq J^Z_{mn}$. (b) For $u_{\uparrow \downarrow}=0.5\omega$ the $B_n^{u_{\uparrow \downarrow}}$ terms do not grow with particle number.
}
\end{centering}
\end{figure*}

For a linear gradient, the direct interaction integrals are not symmetrical under mode exchange: $J^Z_{nm} \neq J^Z_{mn}$. The spin model Hamiltonian includes terms $\hat{H}^{as}=\frac{u_{\uparrow\downarrow}}{8}\sum_{n\neq m} \left(J^Z_{nm}-J^Z_{mn}\right)\left(\hat\sigma^Z_n\hat N_m-\hat\sigma^Z_m\hat N_n\right)$, where $\hat N_n = \hat N^\uparrow_n+\hat N^\downarrow_n$ and $\hat N ^\alpha _n = \hat c^\dagger_{n\alpha} \hat c_{n\alpha}$. These terms, when summed over the index $m$, represent an inhomogeneous magnetic field: $\sum_{m} \hat H^{as} = \sum_nB_n^{u_{\uparrow \downarrow}}\hat \sigma^Z_n$. This combines with the single particle field $B^{sp}_n=\Delta\omega(n+1/2)$ to yield a total  $\hat\sigma^Z_n$ field $B_n\hat \sigma^Z_n$, where $B_n = B^{sp}_n+B_n^{u_{\uparrow \downarrow}}$. We find that even for relatively strong interactions ($u_{\uparrow \downarrow}=0.5\omega$) $B_n^{u_{\uparrow \downarrow}} \ll B^{sp}_n$ for all $n$, as illustrated in Fig.~\ref{zfield}, so these additional terms can be neglected. Additionally, $B_n^{u_{\uparrow \downarrow}}$ does not grow with particle number. Although these terms are not essential for the large-scale features of the dynamics, for completeness we include them in numerical simulations.

\section{Behavior of the Dicke Gap $G$}

We will now discuss the behavior of the gap between the spin-$N/2$ (``Dicke states") and spin-$(N/2-1)$ (``spin wave states"), referred to as the Dicke gap $G$ in the main text, for a general Heisenberg model of the form
$
{\hat{H}=-\frac{1}{4}\sum_{q\ne q'} J_{qq'}\hat{\vec{\sigma}}_q\cdot \hat{\vec{\sigma}}_{q'}.}
$
The only condition we impose is that the coupling matrix $\mathbb{J}$ is real, for compactness of the resulting formulas, and because all couplings considered in this work are real.  Noting that the diagonal terms of $\mathbb{J}$ only contribute an overall constant to the energy and hence do not affect the Dicke gap, they can be ignored.  By direct calculation, the energy of the (degenerate) Dicke states, defined as
${
%|N/2,n_{\mathrm{up}}\rangle=\sqrt{\binom{N}{n_{\mathrm{up}}}^{-1}}\left(\sum_{i=1}^{N} \hat{\mathcal S}^+_i\right)^{n_{\mathrm{up}}}|\downarrow \dots \downarrow\rangle},
|N/2,m_z\rangle=\sqrt{\binom{N}{\frac{N}{2}+m_z}^{-1}}\left(\sum_{i=1}^{N} \hat S^+_i\right)^{\frac{N}{2}+m_z}|\downarrow \dots \downarrow\rangle},
$
with $m_z$ the magnetization, is $E_{\mathrm{Dicke}}=\langle N/2,m_z|-\frac{1}{4}\sum_{q\ne q'} J_{qq'}\hat{\vec{\sigma}}_q\cdot \hat{\vec{\sigma}}_{q'}|N/2,m_z\rangle=-\sum_{q\ne q'}J_{q,q'}/4$.  Because of the SU(2) spin-rotation symmetry of $\hat{H}$, the Dicke states are guaranteed to be eigenstates.  The spin-wave states, which span the total spin-$(N/2-1)$ manifold, can be defined in terms of the Dicke states as 
%|N/2-1,n_{\mathrm{up}},k\rangle&=\sqrt{\frac{\left(N-1\right)}{\left(N-n_{\mathrm{up}}+1\right)\left(N-n_{\mathrm{up}}\right)}}\sum_{n=1}^{N}e^{2\pi i k n/N}\hat{\mathcal S}^+_n|N/2,n_{\mathrm{up}}-1\rangle\, ,
$|N/2-1,m_z,k\rangle=\sqrt{\frac{\left(N-1\right)}{\left(\frac{N}{2}-m_z+1\right)\left(\frac{N}{2}-m_z\right)}}\sum_{n=1}^{N}e^{2\pi i k n/N}\hat S^+_n|N/2,m_z-1\rangle\, ,$
where $k=1,\dots, N-1$.  In the case of a translationally invariant Heisenberg coupling $J_{q,q'}=J_{|q-q'|}$ with $|q-q'|$ the chordal distance, the spin wave states as stated are eigenstates of $\hat{H}$, but when the interactions are not translationally invariant (as is the case for the spin models discussed in this work), the spin wave states only form a basis for the spin-($N/2-1$) subspace.  Straightforward calculations lead to the matrix elements of the Hamiltonian in this subspace:
\begin{align}
%&\langle N/2-1, m_z,k|-\sum_{q\ne q'} J_{q,q'} \hat{\vec{\mathcal{S}}}_q\cdot \hat{\vec{\mathcal{S}}}_{q'}|N/2-1,m_z,k'\rangle =\nonumber\\
%&=\delta_{k,k'}E_{\mathrm{Dicke}}+\frac{1}{N}\sum_{q\ne q'}e^{2\pi i\left(k'-k\right)q/N}J_{qq'}-\frac{1}{N}\sum_{q\ne q'}e^{2\pi i \left(k'q'-k q\right)/N}J_{q,q'}\, .
\langle \frac{N}{2}-1, m_z,k|-\frac{1}{4}\sum_{q\ne q'} J_{q,q'} \hat{\vec{\sigma}}_q\cdot \hat{\vec{\sigma}}_{q'}|\frac{N}{2}-1,m_z,k'\rangle &=\delta_{k,k'}E_{\mathrm{Dicke}}+\frac{1}{N}\sum_{q\ne q'}J_{q,q'}\left[e^{\frac{2\pi i}{N}\left(k'-k\right)q}-e^{\frac{2\pi i}{N} \left(k'q'-k q\right)}\right]\, . \label{gapeqn}
\end{align}
The Dicke gap is then defined as the difference between the smallest eigenvalue of this matrix and the energy of the Dicke states.  As two concrete examples, in the all-to-all case, $J_{q,q'}=J\left(1-\delta_{q,q'}\right)$, the Dicke gap is $G=JN$, and in the nearest-neighbor case $J_{q,q'}=\delta_{\left|q-q'\right|,1}J$, $G=2\left(1-\cos\left(2\pi/N\right)\right)\sim \frac{4\pi^2}{N^2}+\mathcal{O}\left(1/N^3\right)$.  These examples illustrate the general observation that long-range, near-collective interactions cause the Dicke gap to grow with particle number, while the Dicke gap decreases with $N$ for sufficiently short-range interactions.

\begin{figure*} %Retain asterik for wide figure and wide figure captions
\begin{centering}
\includegraphics[width=0.9\textwidth]{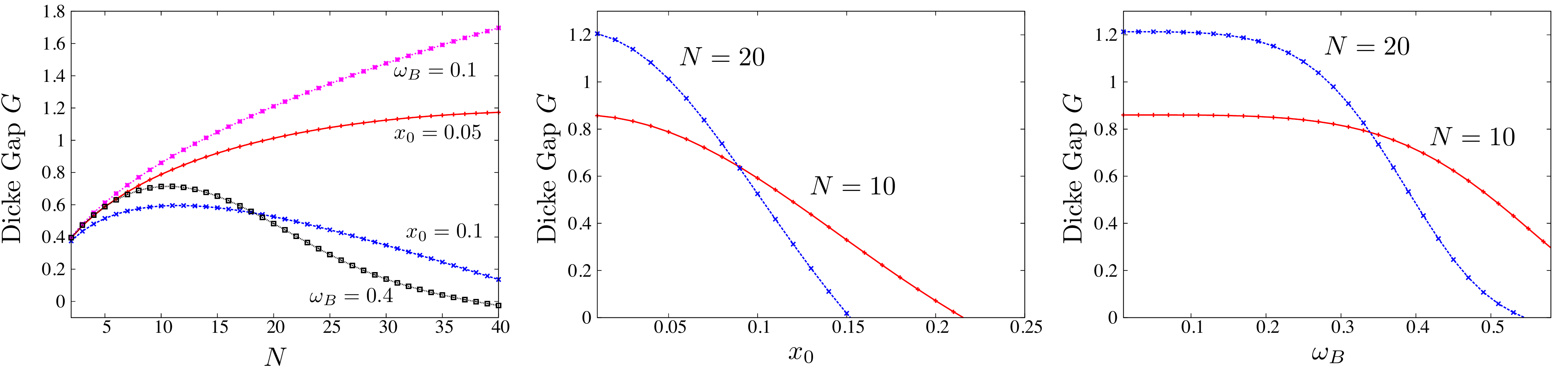}
\caption
{ \label{fig:DGap}
\raggedright
Scaling of the Dicke gap $G$ with magnetic field gradient and particle number. (Left panel) Scaling of the Dicke gap with particle number for constant gradients of strength $x_0=0.05$ and $0.1$ and linear gradients of strength $\omega_B=0.1$ and $0.4$ (all quantities are measured in oscillator units).  All gaps increase with $N$ to a certain gradient-dependent critical value and then decrease, with larger gaps for smaller gradients. (Right panels) Scaling of the Dicke gap at fixed particle number $N=10$ and 20 with the constant (center panel) or linear (right panel) gradient strength.  The gap closes for weaker gradient as the particle number increases, demonstrating that larger particle numbers require smaller gradients to be in the near-Heisenberg regime.  For even smaller gradients, the gap increases with particle number, more effectively enforcing collective behavior.
}
\end{centering}
\end{figure*}

In Fig.~\ref{fig:DGap} we show the Dicke gap $G$ for a Heisenberg model with $J_{q,q'}=J^{\perp}_{q,q'}$, where $\mathbb{J}^{\perp}$ corresponds to different realizations of the energy-lattice spin model.  The leftmost panel shows the scaling of the gaps at fixed gradient strength with particle number.  The gaps are always larger for smaller gradient strength, showing that smaller gradients always lead to a more collective, near-Heisenberg behavior.  The rightmost panels show the behavior of the gaps at fixed particle number as a function of gradient strength.  For any fixed number of particles, there is a finite critical gradient strength where the Dicke gap closes.  This critical gradient decreases with increasing particle number.  However, at small enough gradient strengths, the Dicke gap is larger for increasing particle number.  This demonstrates that increasing the particle number can either increase or decrease the Dicke gap.

\section{ Spin segregation}
To understand spin segregation in a many body system we have to consider the coupling of the Dicke states $|N/2,m_z\rangle$ to sectors with different total $S$. To first order, local spin operators $\hat \sigma^\alpha _n$ couple the Dicke states to the spin wave states $|N/2-1,m_z,k\rangle$~\cite{Swallows2011}.
We will examine the dynamics within this subspace, assuming the population in the spin wave sector is much smaller than that of the Dicke sector, suppressed by the small parameter $\Delta \omega/u_{\uparrow \downarrow}$. We also assume that the interactions are fully collective for simplicity. The state of the system can be written as
\begin{eqnarray}
|\psi\rangle=\displaystyle\sum\limits_{m}c_m|m\rangle+\displaystyle\sum\limits_{m,k}d_{mk}|mk\rangle,
\end{eqnarray}
where $|m\rangle$ are the Dicke states, labeled by their magnetization $m=-N/2, \cdots N/2-1, N/2$, $N$ is the total number of particles, and $|mk\rangle$ are the spin wave states $|N/2-1,m_z,k\rangle$ where  $k = 1, \cdots, N-1$.
It is useful to define the matrix elements \cite{Swallows2011}
\begin{eqnarray}
&&\langle m | \hat \sigma^Z_n |m' \rangle = \frac{2m}{N}\delta_{mm'} \nonumber\\
&&M^n_{mm'k} = \langle m | \hat \sigma^Z_n | m'k \rangle = 2e^{2\pi i kn/N}\sqrt{\frac{(N/2)^2-m^2}{N^2(N-1)}}\delta_{m,m'}\sim \frac{1}{\sqrt{N}}, \nonumber \\
&&M^n_{mk} \equiv M^n_{mmk}, \nonumber \\
&&P^n_{mkm'k'}=\langle mk | \hat \sigma^Z_n | m'k' \rangle=\left(-2e^{2\pi i (k'-k)n/N}+N\delta_{k,k'}\right)\frac{2m}{N(N-2)}\delta_{mm'}\sim \frac{1}{N}.
\end{eqnarray}
(Note that ``$n$" on the matrix elements is a superscript and not a power.) The $M$ and $P$ matrix elements scale differently with $N$ such that the $M$ elements will dominate in the thermodynamic limit. 

We take the Heisenberg (weak gradient) limit of the interaction Hamiltonian combined with the single particle Hamiltonian which contains inhomogeneous terms $n\Delta\omega\hat\sigma^Z_n$ which induce transitions outside of the Dicke Manifold:
\begin{eqnarray}
\hat H=-\frac{u_{\uparrow \downarrow}}{4}\sum_{n \neq m}J^\perp_{nm} \vec{\sigma}_n\cdot\vec{\sigma}_m + \displaystyle\sum\limits_n \left[\bar\omega(n + 1/2)\hat N_n + \Delta\omega \left(n+1/2\right)\hat \sigma^Z_n \right],
\end{eqnarray}
We assume all the spin wave states have zero energy and the Dicke manifold is offset by the Dicke gap $G$. In the basis of Dicke and spin wave states the Hamiltonian is
\begin{eqnarray}
\hat H = E^0_{\bar{n}}-G\displaystyle\sum\limits_{m}|m\rangle\langle m| + \displaystyle\sum\limits_n \Delta\omega\left(n+1/2\right)\displaystyle\sum\limits_{m,m',k,k'}\left(M^n_{m k}|m\rangle\langle mk |+P^n_{mkm'k'}|mk\rangle\langle m'k'| + {\rm H.c.}\right).
\end{eqnarray}
Where $E^0_{\bar{n}}=N\bar{\omega}\left(\bar{n}+1/2\right)$, which depends on the set of occupied modes and will contribute an additional dynamical phase to $|\psi\rangle$ which will not contribute to the dynamics. We can use the fact that $M^n_{mk} \gg P^n_{mkm'k'}$ for $N \gg 1$ and drop the $P^n_{mkm'k'}$  terms. The Schrodinger equation implies
\begin{eqnarray}
&&i \dot{c}_m=-Gc_m+\displaystyle\sum\limits_{n}\Delta\omega\left(n+1/2\right)M^n_{mk}d_{mk} \nonumber \\
&& i \dot{d}_{mk} = \displaystyle\sum\limits_{n}\Delta\omega\left(n+1/2\right)M^{n*}_{mk}c_m.
\end{eqnarray}
Assuming the population stays mostly in the Dicke manifold implies $c_m \gg d_{mk}$. Using this and $\Delta\omega \ll 1$ the equation of motion for $c_m$ can thus be approximated as $i\dot{c}_m = -Gc_m$. With this additional approximation,
\begin{eqnarray}
&&c_m(t)=c_m(0)e^{iGt} \nonumber \\
&&d_{mk}(t)=\displaystyle\sum\limits_{n}\Delta\omega\left(n+1/2\right)\frac{c_m(0)M^{n*}_{mk}}{G}\left(1-e^{iGt}\right),
\end{eqnarray}
where for a spin polarized sample initially pointing in the $X$-direction, the Dicke state coefficients are
\begin{eqnarray}
c_m(0)=\sqrt{\frac{1}{2^N}\binom{N}{\frac{N}{2}+m}}.
\end{eqnarray}
The expectation of a generic spin operator is
\begin{eqnarray}
&&\langle \hat S^\alpha \rangle = \displaystyle\sum\limits_{m,m'} c^*_mc_{m'}\langle m | \hat S^\alpha | m'\rangle+\displaystyle\sum\limits_{m,k,m'} d^*_{mk}c_{m'}\langle mk | \hat S^\alpha | m'\rangle+\nonumber\\
&&+\displaystyle\sum\limits_{m,m',k'} c^*_{m}d_{m'k'}\langle m | \hat S^\alpha | m'k'\rangle + \displaystyle\sum\limits_{m,k,m',k'} d^*_{mk}d_{m'k'}\langle mk | \hat S^\alpha | m'k'\rangle.
\end{eqnarray}
Note that $d^*_{mk}d_{m'k'} \ll c^*_{m}d_{m'k'}$ so we ignore those terms. In our case we have
\begin{eqnarray}
&&\langle \hat \sigma^Z_{n'} \rangle =\frac{4\Delta\omega}{2^{N}N^2(N-1)G}\displaystyle\sum\limits_{n,m,k}\left(n+1/2\right)\binom{N}{\frac{N}{2}+m}\left((N/2)^2-m^2\right)e^{2\pi i k(n-n')/N}\left(e^{iGt}-1\right) +{\rm H.c.}
 \nonumber \\
&&=\frac{2\Delta\omega}{G}\left(n'-N_{{\bf n}^i}^{{\rm ave}}\right)\left(\cos\left(Gt\right)-1\right). \label{sigmazn}
\end{eqnarray}
where $N_{{\bf n}^i}^{{\rm ave}}$ is the average mode number of the set of occupied modes ${\bf n}^i$. (In the above derivation the spin label $n$ was arbitrary and the results hold for any configuration ${\bf n}^i$ of $N$ total spins.) Notice that the dynamics of $\sigma^Z_{n'}$ depends linearly on $n'$ and changes sign when $n' > N_{{\bf n}^i}^{{\rm ave}}$: high energy modes evolve differently from low energy modes, which is the origin of spin segregation.

\section{Scaling of dynamical quantities}
\begin{figure*} %Retain asterik for wide figure and wide figure captions
\begin{centering}
\includegraphics[scale=.5]{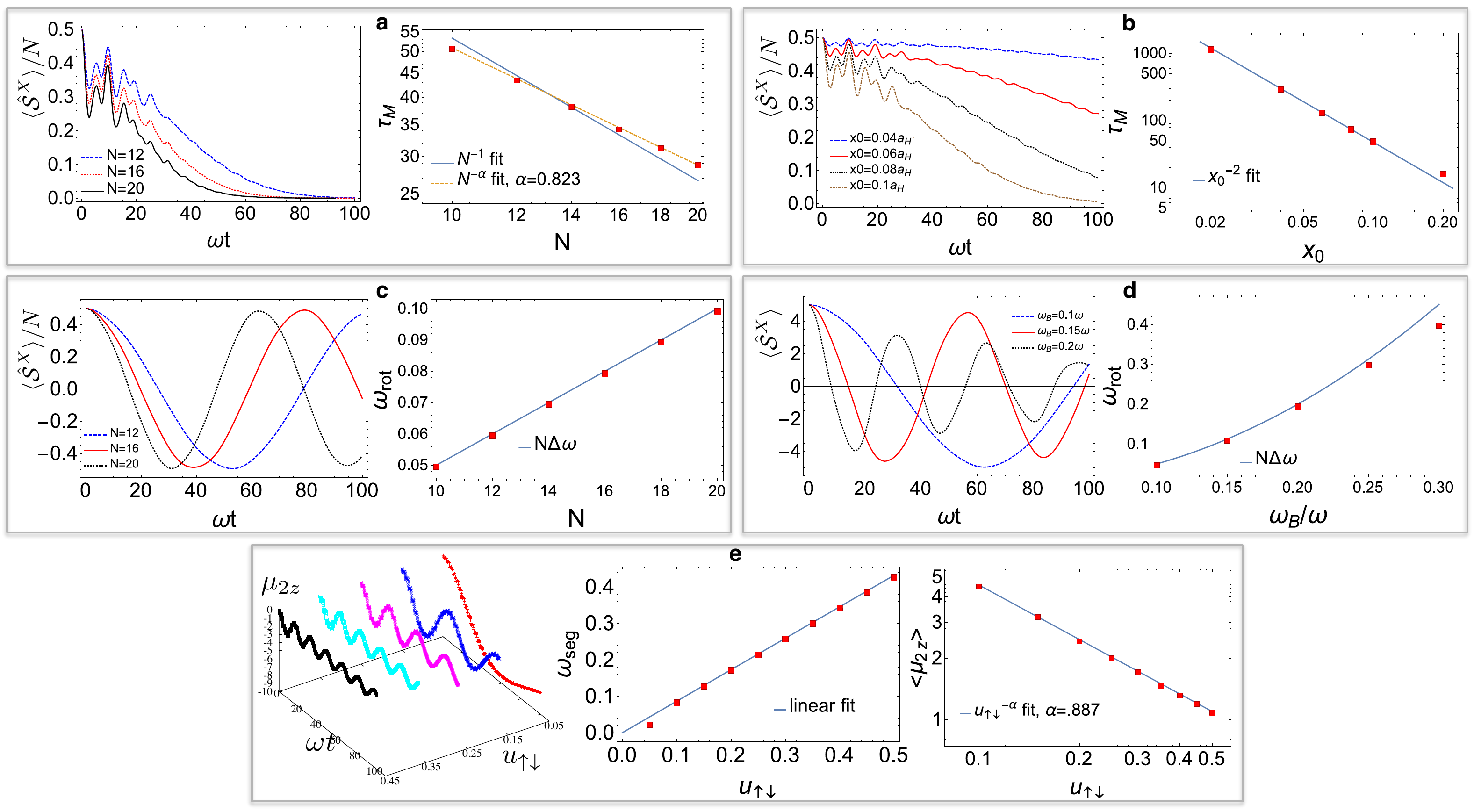}
\caption
{ \label{const_scaling}
\raggedright
Scaling. (a) Dynamics vs. N for a constant gradient $x_0=0.1a_H$, from which $\tau_M$ is extracted and found to scale like $\tau_M \sim N^{-.823}$, close to the $N^{-1}$ prediction. (b) Dynamics and scaling of $\tau_M$ vs. $x_0$ which agrees well with the prediction $x_0^{-2}$, for $N=10$. (c) $\langle \hat{\mathcal S}^X \rangle /N$ vs. $N$, when $\omega_B = 0.1\omega$, from which $\omega_{rot}$ is extracted and agrees well with the prediction $\omega_{rot} = N\Delta\omega$. (d) $\langle  \hat{\mathcal S}^X \rangle$ vs. $\omega_B$ for $N=10$. Predictions fail when $\omega_B \sim u_{\uparrow \downarrow}$. (All cases are $u_{\uparrow \downarrow} = 0.35\omega$.) (e) $\mu_{2z}$, vs. $u_{\uparrow \downarrow}$; oscillations become more pronounced for stronger interactions. $\omega_{seg}$ scales linearly with $u_{\uparrow \downarrow}$. $\langle \mu_{2z}\rangle \sim u_{\uparrow \downarrow}^{-.887}$, close to the prediction of $u_{\uparrow \downarrow}^{-1}$.}
\end{centering}
\end{figure*}

The short time dynamics of a generic XXZ Hamiltonian for a state initially polarized along the $X$ direction is~\cite{Hazzard2014}
\begin{eqnarray}
\langle \hat{S}^X \rangle= \frac{N}{2} - \frac{(u_{\uparrow \downarrow} t)^2}{16}\displaystyle\sum\limits_{n \neq m} \Delta_{nm}^2 + O(t^3) \approx \langle \hat{S}^X \rangle_{t=0} \left(1-(t/\tau_M)^2\right), \quad \tau_M =\frac{1}{u_{\uparrow\downarrow}}\sqrt{\frac{2N}{\displaystyle\sum\limits_{n \neq m} \Delta_{nm}^2}}, \label{stdynamics}
\end{eqnarray}
where $\Delta_{nm} \equiv J^Z_{nm}-J^\perp_{nm}$ and $\tau_M$ is defined as the demagnetization time. 
For a linear gradient we expand the parameters in $x_0/a_H$, set $a_H=1$, and find
$\Delta_{nm} =  J^Z_{nm}-J^\perp_{nm} \approx 2x_0^2\Lambda_{nm}$, where
\begin{eqnarray}
&&\Delta_{nm} \approx 
n J^0_{n-1,m}-2 \sqrt{nm} J^0_{n-1,n,m-1,m}-2 \sqrt{n(n+1)} J^0_{n-1,n+1,m,m} + \nonumber \\
&&-2 \sqrt{m(m+1)} J^0_{n,n,m-1,m+1}+(1+m)J^0_{n,m+1}+ mJ^0_{n,m-1}-2 \sqrt{(m+1)(n+1)} J^0_{n,n+1,m,m+1} + \nonumber\\
&& +(1+n)J^0_{n+1,m}+2 \sqrt{n(m+1)} J^0_{n-1,n,m,m+1}+2 \sqrt{m(n+1)} J^0_{n,n+1,m-1,m}, \label{Lambda}
\end{eqnarray}
$J^0_{nmpq}=\int_{-\infty}^\infty dx\phi_n(x)\phi_m(x)\phi_p(x)\phi_q(x)$, and $J^0_{nm} \equiv J^0_{nnmm}$. The formula
$
\overline{\Lambda_{nm}} \approx \overline{nJ^0_{nm}} \sim \sqrt{N}
$ works well, where $\overline{X_{nm}} \equiv \sum_{{n,m}\in{\bf n}^i} X_{nm}/(N(N-1))$ is the arithmetic average and we have used $\overline{J^0_{nm}} \sim 1/\sqrt{N}$. We find that for $x_0 \ll a_H$, $\overline{\Delta_{nm}} \sim x_0^2\sqrt{N}$. Further assuming $\overline{ \Delta_{nm}^2} \approx (\overline{\Delta_{nm}})^2$, this implies $\tau_M \sim \left(Nu_{\uparrow \downarrow}x_0^2\right)^{-1}$. Fig.~\ref{const_scaling}(a) shows the scaling of $\tau_N$ vs. $N$. Fitting the dynamics to a Gaussian decay function $A\exp(-t^2/\tau_M^2)$ we find that $\tau_M\sim N^{-.823}$, close to the prediction of $N^{-1}$. In Fig.~\ref{const_scaling}(b) we show the scaling of $\tau_M$ vs. $x_0$, which agrees well with the $x_0^{-2}$ prediction.  

In Fig.~\ref{const_scaling}(c,d) we show how $\langle  \hat{\mathcal S}^X \rangle$ depends on $N$ and $\omega_B$, respectively. We use a cosine fitting function $A\cos(\omega_{rot}t)$ to extract the collective Bloch vector precession frequency $\omega_{rot}$ and compare with the prediction $N\Delta\omega$. In Fig.~\ref{const_scaling}(c) we use $\omega_B=0.1\omega$ and a relatively large interaction strength $u_{\uparrow \downarrow} = 0.35\omega \gg \Delta\omega$. This is the self-rephasing regime so the prediction works well. In Fig.~\ref{const_scaling}(d) we fix $N=10$ and $u_{\uparrow \downarrow} = 0.35\omega$, and vary $\omega_B$. We see deviations from the prediction for large $\omega_B$, because interactions are not strong enough to protect against population leakage outside of the Dicke manifold. 

We can quantify spin segregation by the second moment of the spin density $\mu_{2z} = 2\int_{-\infty}^\infty dx x^2\langle \hat{\mathcal{S}}^Z(x) \rangle$. For a homogeneous spin distribution, $\mu_{2z}=0$. When the $\uparrow$ ($\downarrow$) spins are concentrated more towards the edges of the trap, the sign of $\mu_{2z}$ is positive (negative). In Fig.~\ref{const_scaling}(e) we plot $\mu_{2z}$ dynamics for various interaction strengths, fixing $N=10$ and $\omega_B=0.1\omega$. For larger interactions the oscillations become smaller, faster, and less damped, confirming the ``Rabi oscillation" behavior of spin segregation. We fit $\mu_{2z}$ to an offset cosine function $A + B\cos(\omega_{seg}t+\phi)$ to extract the scaling of the segregation frequency $\omega_{seg}$, and the average value of the segregation, $\langle \mu_{2z} \rangle = A$. A linear fit of $\omega_{seg}$ vs. $u_{\uparrow \downarrow}$ with slope of 0.86 confirms linear scaling with interaction energy. We find $\langle \mu_{2z} \rangle \sim u_{\uparrow \downarrow}^{-0.887}$, close to the prediction of $u_{\uparrow \downarrow}^{-1}$. 

\section{Comparison with experiment in Ref.~\cite{Du2008}}
To make a comparison with experiment we first benchmarked the system with a one dimensional DMRG simulation of the dynamics to determine the role of single particle motion in the experiment. In this regime DMRG is fully reliable. The experiment in Ref.~\cite{Du2008} was conducted with $2\times10^5$ atoms in a cigar-shaped geometry with trapping frequencies $\{\omega_x, \omega_y, \omega_z\}=\{145\times2\pi ~{\rm Hz}, 4360\times2\pi ~{\rm Hz}, 4360\times2\pi~ {\rm Hz}\}$. A zero temperature version of this gas would fill up the harmonic oscillator modes in the lowest energy configuration, resulting in about $560$ particles in the $x$-direction (occupying modes $n_x=0$ through $n_x=559$) and $19$-particles in each of the transverse directions. Our simulation used $N=560$, with $n_x=n_y=0$ for all the particles, and the results are shown in Fig.~3(b,c) of the main text. From this simulation we concluded that coherences between mode sectors are unimportant since single particle motion is negligible. The lack of single particle motion is due to the very small inhomogeneity along the $x$-direction: $\Delta\omega = \left(\omega^\uparrow - \omega^\downarrow\right)/2\omega_x = 8.62\times10^{-6}$.
\begin{figure}[h]
	\centering
	\includegraphics[width=400pt]{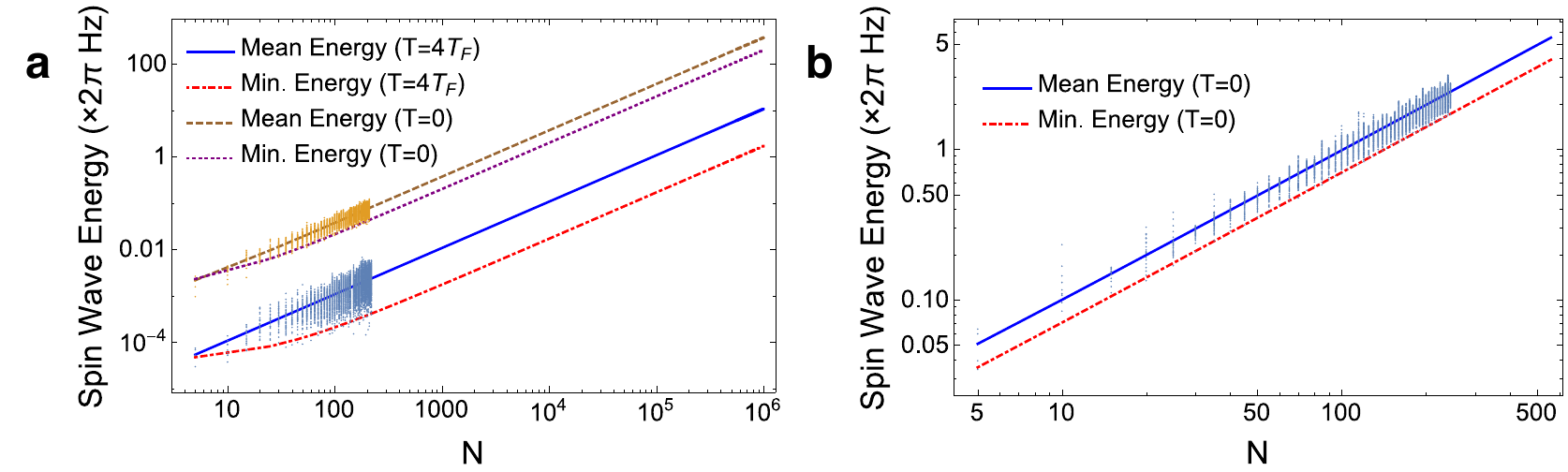}
	\caption{(a) Spin wave energies vs. particle number for a 3D system with parameters taken from \cite{Du2008}, based on Monte Carlo sampling of harmonic oscillator mode configurations. Extrapolated to $N=2\times10^5$ particles, at $T=4T_F$ the average energy is $\approx 2.12\times2\pi ~{\rm Hz}$ and the Dicke gap (minimum energy) is $\approx 0.34\times2\pi ~{\rm Hz}$, both much smaller than the single particle inhomogeneity of $4.85\times2\pi~{\rm Hz}$. For $N=2\times10^5$ particles at $T=0$ the Dicke gap is $\approx 39.53 \times 2\pi~{\rm Hz}$. (b) Spin wave energies vs. particle number for a 1D system at $T=0$ with parameters taken from \cite{Du2008}. At $N=560$ the Dicke gap is $3.92\times2\pi~{\rm Hz}$, much larger than the inhomogeneity of $0.70\times2\pi ~{\rm Hz}$. \label{gaps}} 
\end{figure}

The experiment was conducted at a high temperature of $27\mu {\rm K} \approx 4T_F$, where $T_F$ is the Fermi temperature. The average harmonic oscillator mode occupations were: $\bar N_i \approx \hbar\omega_i/k_BT : \{\bar N_x,\bar N_y ,\bar N_z \}=\{3878,129,129\}$. We performed a Monte Carlo sampling of the energy separation of the spin-wave and Dicke states, where mode configurations were sampled from a thermal distribution and the spin wave energies were computed and plotted in Fig.~\ref{gaps}(a). The mean, standard deviation, and minimum (Dicke gap) of the spin wave energies increases linearly with particle number, allowing us to extrapolate to higher particle number. For $2\times10^5$ particles at $T=4T_F$ the Dicke gap is $\approx 0.34\times2\pi ~{\rm Hz}$, the average energy of the spin wave states is $\approx 2.12\times2\pi ~{\rm Hz}$, and the standard deviation of the energies is $\approx 1.13\times2\pi ~{\rm Hz}$. A typical magnitude of the coupling via the inhomogeneity is $\bar N_x\Delta\omega = 4.85\times2\pi~{\rm Hz}$, much larger than all of these energies. The typical thermal energy per particle is also much higher than all of these energies.

In such a high temperature system the protection from the Dicke gap is significantly suppressed and the long time dynamics are potentially difficult to analyze. However in Ref.~\cite{Du2008} the spin density at the cloud center, ${(n^\uparrow(x=0)-n^\downarrow(x=0))/n_0}$, exhibited a damped oscillation that quickly reached an asymptotic value of ${(n^\uparrow(x=0)-n^\downarrow(x=0))/n_0|_{t\rightarrow\infty} \equiv \overline{\Delta n}}$. Since initially all the atoms were prepared in the Dicke manifold, the initial transfer of population from the Dicke manifold to the spin wave manifold that happens at short times should be captured by our analytic expressions. To match the short time to the long time dynamics we use the asymptotic value of the population, $\overline{\Delta n}$ as a fitting parameter. We compute the thermal average by sampling our analytic expression over a Gaussian distribution of Dicke gaps. The mean, $G_0$, and the standard deviation, $\Delta G$, were extracted by a Monte-Carlo sampling of the gaps  evaluated from matrices constructed accordingly to Eq.~\ref{gapeqn}. In the limit of a sum of many such oscillations, the dynamics can be approximated as an integral:
\begin{eqnarray}
(n^\uparrow(0)-n^\downarrow(0))/n_0 \approx \overline{\Delta n} \int dG \left(1-\cos(Gt)\right)\frac{e^{-\frac{(G-G_0)^2}{2\Delta G^2}}}{\sqrt{2\pi}\Delta G}=\overline{\Delta n}\left(1-\cos(G_0t)\right)e^{-\frac{\left(\Delta G t\right)^2}{2}}. \label{cloudcenter}
\end{eqnarray}
The thermal average of the population imbalance extracted from Eq.~\ref{cloudcenter} agrees well with the data from \cite{Du2008} and is shown in Fig.~3(d) of the main text.

\section{ Matrix product state simulations}
The variational matrix product state (MPS) studies of the main text were performed using extensions of the open source MPS library~\cite{OSMPS,Wall_Carr_12}.  We use an MPS ansatz which explicitly conserves total particle number, but does not conserve the total magnetization.  While the dynamics preserve the total magnetization, the initial collective rotation of spins along the $x$ direction involves a sum over many different magnetization sectors, and so leaving the magnetization unconstrained is convenient.  Following this collective rotation, the next step is to enact the sudden quench of trapping parameters, which amounts to applying a spin-dependent displacement ($\psi\left(x\right)\to \psi\left(x+\lambda\right)$, constant gradient) or spin-dependent dilation ($\psi\left(x\right)\to \sqrt{\lambda}\psi\left(\lambda x\right)$, linear gradient) to the single-particle states.  Since we assume harmonic traps, the displacement and dilation operators are known analytically as
\begin{align}
\hat{U}_{\mathrm{displacement}}&=e^{\left(\hat{a}-\hat{a}^{\dagger}\right) \lambda/(\sqrt{2} a_H)}\, , \nonumber \\
\hat{U}_{\mathrm{dilation}}&=e^{\ln\lambda \left(\hat{a}^2-(\hat{a}^{\dagger})^2\right)/2}\, ,
\end{align}
where $\hat{a}$ and $\hat{a}^{\dagger}$ are the ladder operators of the original (no gradient) harmonic oscillator.  Writing these ladder operators in second quantized form on the energy lattice, the basis transformations above take the form of time evolution under a hopping model with spin-dependent and inhomogeneous hopping amplitudes.  Here, time evolution refers to the fact that the operation consists of applying the exponential of an anti-Hermitian many-body operator.  In the constant gradient case, the hopping model contains only nearest-neighbor hopping, while the linear gradient case is a model with only next-nearest neighbor hopping.  We enact this effective time evolution by decomposing it into a product of few-site unitaries using a Trotter decomposition with the error controlled by a small ``step size" $\Delta \lambda$, and then applying these few-site unitaries to the MPS via standard techniques~\cite{Schollwoeck}.

Next, we wish to perform time evolution under the long-range spin model
\begin{eqnarray}
&&\hat H=
\frac{u_{\uparrow\downarrow}}{4}\displaystyle\sum\limits_{{n \neq m }}\left[J^Z_{nm}\left(\hat{N}_n  \hat{N}_m - \hat{\sigma}^Z_n  \hat{\sigma}^Z_m\right)-J^\perp_{nm}\left(\hat{\sigma}^X_n  \hat{\sigma}^X_m + \hat{\sigma}^Y_n  \hat{\sigma}^Y_m\right)+\frac{1}{2} \left(J^Z_{nm}-J^Z_{mn}\right)\left(\sigma^Z_n\hat N_m-\sigma^Z_m\hat N_n\right)\right]\nonumber\\
\label{eq:FullHspinMo}&&+u_{\uparrow\downarrow}\displaystyle\sum\limits_{{n }}J_{nn} \hat N_n^\uparrow \hat N_n ^\downarrow+  \displaystyle\sum\limits_n \left[\bar\omega(n + 1/2)\hat N_n + \Delta\omega \left(n+1/2\right)\hat \sigma^Z_n \right],
\end{eqnarray}
where $J_{nn}  \equiv A_{nnnn}$.  We perform time evolution using the second-order method of Zaletel \emph{et al.}~\cite{Zaletel}.  In this method, an explicit matrix product operator (MPO) approximation to the propagator $\hat{U}$ is formed from the MPO form of the Hamiltonian, which is then applied to the state at time $t$, $|\psi\left(t\right)\rangle$ by variational minimization of the functional $\left||\phi\rangle-\hat{U}|\psi\left(t\right)\rangle\right|^2$ over all MPSs $|\phi\rangle$ with fixed resources.  For the variational minimization, we perform four sweeps per timestep and impose an upper limit on the discarded weight per bond of $10^{-9}$.  The maximum bond dimension used in the simulations of this work is roughly 2000.

In order to apply the method of Zaletel \emph{et al.}, we must construct an MPO representation of the Hamiltonian Eq.~\eqref{eq:FullHspinMo}.  For long-range interactions which are translationally invariant, $\hat{H}=\sum_{i<j} f\left(j-i\right)\hat{A}_i \hat{B}_j$, a well-established procedure exists for converting this interaction into an MPO~\cite{Crosswhite, Murg}.  In this procedure, the function $f\left(r\right)$ is fitted to a sum of $n_{\mathrm{exp}}$ exponentials via the ansatz $\tilde{f}\left(r\right)=\sum_{n=1}^{n_{\mathrm{exp}}} J_n \lambda_n^{r}$, and then a known MPO construction of exponentially decaying interactions is used.  Interactions on the single-particle mode space lattice are not translationally invariant, and so this procedure does not apply.  However, we have devised a related procedure, in which an inhomogeneous interaction $\hat{H}=\sum_{i<j} f\left(i,j\right)\hat{A}_i \hat{B}_j$ is modeled by a sum of exponentials with site-dependent weights and exponential decay parameters via the ansatz $\tilde{f}\left(i,j\right)=\sum_{n=1}^{n_{\mathrm{exp}}} J_{i,n}\prod_{k=i}^{j-1}\lambda_{k,n}$.  These parameters are variationally optimized using an alternating least squares algorithm.  Imposing the condition that the residual $\sum_{i<j}\left|f\left(i,j\right)-\tilde{f}\left(i,j\right)\right|^2<10^{-7}$ leads to approximations with $ n_{\mathrm{exp}}\sim 7$ exponentials.

\end{document}